\documentclass[traditabstract, longauth]{aa}  

\usepackage{booktabs}
\usepackage{graphicx}
\usepackage{txfonts}
\usepackage{multirow}
\usepackage{tabularx}
\usepackage{hyperref}
\hypersetup{colorlinks,citecolor=blue}
\usepackage{amstext}

\newcommand{\ha}{H$\alpha$}
\newcommand{\hb}{H$\beta$}
\newcommand{\oiii}{[O\,\textsc{iii}]}
\newcommand{\nii}{[N\,\textsc{ii}]}

\newcommand{\IAA}{1}
\newcommand{\CEFCA}{2}
\newcommand{\CEFCAA}{3}
\newcommand{\IFAE}{4}
\newcommand{\USP}{5}
\newcommand{\UNIVIE}{6}
\newcommand{\IFCA}{7}
\newcommand{\LAM}{8}
\newcommand{\FAPESP}{9}
\newcommand{\NOIRLAB}{10}
\newcommand{\ON}{11}
\newcommand{\UFES}{12}
\newcommand{\INAFTS}{13}
\newcommand{\IFPU}{14}
\newcommand{\DIPC}{15}
\newcommand{\IAC}{16}
\newcommand{\ULL}{17}
\newcommand{\CNAO}{18}
\newcommand{\IAGUSP}{19}
\newcommand{\INSTRUMENTSFOUR}{20}

\begin{document}

   \title{OJALÁ: Optimizing J-PAS Astronomy for Large-scale Analysis} 
   \subtitle{A transformer-based model for source classification and physical-parameter inference}

   \titlerunning{OJAL\'A: source classification and parameter inference}
   \authorrunning{Mart\'inez-Solaeche et al.}


    \author{
    G.~Mart\'inez-Solaeche \inst{\IAA}
    \and
    R.~M.~Gonz\'alez Delgado \inst{\IAA}
    \and
    R.~Garc\'ia-Benito \inst{\IAA}
    \and
    A.~Hern\'an-Caballero \inst{\CEFCA,\CEFCAA}
    \and
    I.~P\'erez-R\`afols \inst{\IFAE}
    \and
    L.~A.~D\'iaz-Garc\'ia \inst{\IAA}
    \and
    L.~Raul~Abramo \inst{\USP}
    \and
    J.~E.~Rodr\'iguez-Mart\'in \inst{\IAA}
    \and
    A.~M.~Conrado \inst{\IAA}
    \and
    I.~Breda \inst{\UNIVIE}
    \and
    H.~Dom\'inguez~S\'anchez \inst{\IFCA}
    \and
    I.~M\'arquez \inst{\IAA}
    \and
    M.~Pieri \inst{\LAM}
    \and
    D.~L\'opez-Cano \inst{\USP,\FAPESP}
    \and
    V.~M.~Placco \inst{\NOIRLAB}
    \and
    L.~Nakazono \inst{\ON}
    \and
    A.~del~Pino \inst{\IAA}
    \and
    V.~Marra \inst{\UFES,\INAFTS,\IFPU}
    \and
    J.~Alcaniz \inst{\ON}
    \and
    N.~Benitez
    \and
    S.~Bonoli \inst{\DIPC}
    \and
    S.~Carneiro \inst{\ON}
    \and
    A.~J.~Cenarro \inst{\CEFCA,\CEFCAA}
    \and    
    D.~Crist\'obal-Hornillos \inst{\CEFCA}
    \and
    S. Daflon \inst{\ON}
    \and
    R.~A.~Dupke \inst{\ON}
    \and
    A.~Ederoclite \inst{\CEFCA}
    \and
    C.~Hern\'andez-Monteagudo \inst{\IAC,\ULL}
    \and
    J. Liu \inst{\CNAO}
    \and
    C.~L\'opez-Sanjuan \inst{\CEFCA,\CEFCAA}
    \and
    A.~Mar\'in-Franch \inst{\CEFCA,\CEFCAA}
    \and
    C.~Mendes~de~Oliveira \inst{\IAGUSP}
    \and
    M.~Moles \inst{\CEFCA,\IAA}
    \and
    F. Roig \inst{\ON}
    \and
    L.~Sodr\'e \inst{\IAGUSP}
    \and
    K.~Taylor \inst{\INSTRUMENTSFOUR}
    \and
    J.~Varela \inst{\CEFCA}
    \and
    H.~V\'azquez~Rami\'o \inst{\CEFCA,\CEFCAA}
    \and
    J.~M.~V\'ilchez \inst{\IAA}
    \and
    J. Zaragoza-Cardiel \inst{\CEFCA,\CEFCAA}
    }

    \institute{
    Instituto de Astrof\'isica de Andaluc\'ia (CSIC), PO Box 3004, 18080 Granada, Spain\\
    \email{gimarso@iaa.es}
    \and
    Centro de Estudios de F\'isica del Cosmos de Arag\'on (CEFCA), Plaza San Juan, 1, 44001 Teruel, Spain
    \and
    Unidad Asociada CEFCA-IAA, CEFCA, Unidad Asociada al CSIC por el IAA y el IFCA, Plaza San Juan 1, 44001 Teruel, Spain
    \and
    Departament de Física, EEBE, Universitat Politècnica de Catalunya, c/Eduard Maristany 10, 08930 Barcelona, Spain
    \and
    Departamento de F\'isica Matem\'atica, Instituto de F\'isica, Universidade de S\~ao Paulo, Rua do Mat\~ao, 1371, CEP 05508-090, S\~ao Paulo, Brazil
    \and
    Department of Astrophysics, University of Vienna, T\"urkenschanzstra\ss e 17, 1180 Vienna, Austria
    \and
    Instituto de F\'isica de Cantabria, Av.~de los Castros, 39005 Santander, Cantabria, Spain
    \and
    Aix Marseille Univ, CNRS, CNES, LAM, Marseille, France
    \and
    Funda\c{c}\~ao de Amparo \`a Pesquisa do Estado de S\~ao Paulo (FAPESP), R.~Pio XI, 1500, Alto da Lapa, 05468-901, S\~ao Paulo, SP, Brazil
    \and
    NSF NOIRLab, Tucson, AZ 85719, USA
    \and
    Observat\'orio Nacional, Rua General Jos\'e Cristino, 77, S\~ao Crist\'ov\~ao, 20921-400, Rio de Janeiro, Brazil
    \and
    Departamento de F\'isica, Universidade Federal do Esp\'irito Santo, 29075-910, Vit\'oria, ES, Brazil
    \and
    INAF, Osservatorio Astronomico di Trieste, via Tiepolo 11, 34131 Trieste, Italy
    \and
    IFPU, Institute for Fundamental Physics of the Universe, via Beirut 2, 34151 Trieste, Italy
    \and
    Donostia International Physics Center, Paseo Manuel de Lardizabal 4, E-20018 Donostia-San Sebasti\'an, Spain
    \and
    Instituto de Astrof\'isica de Canarias, C/ V\'ia L\'actea, s/n, E-38205, La Laguna, Tenerife, Spain
    \and
    Universidad de La Laguna, Avda Francisco S\'anchez, E-38206, San Crist\'obal de La Laguna, Tenerife, Spain
    \and
    National Astronomical Observatories, Chinese Academy of Sciences, Beijing 100101, PR China
    \and
    Universidade de S\~ao Paulo, Instituto de Astronomia, Geof\'isica e Ci\^encias Atmosf\'ericas, Rua do Mat\~ao 1226, 05508-090, S\~ao Paulo, Brazil
    \and
    Instruments4, 4121 Pembury Place, La Ca\~nada-Flintridge, CA 91011, USA
    }

   \date{}

    \abstract{
    The advent of large-scale surveys requires efficient machine
    learning techniques to exploit the information content of massive
    datasets. We present \texttt{OJALA}, a transformer-based
    autoregressive model designed to simultaneously classify
    astronomical objects and infer their physical parameters using
    54-band photometry from the Javalambre Physics of the Accelerating
    Universe Astrophysical Survey ($\text{J-PAS}$), combined with
    broad-band photometry from the DESI Legacy Imaging Surveys
    and WISE. The model is trained on approximately 20 million
    synthetic SEDs generated from DESI DR1 spectra. We validate
    \texttt{OJALA} using a cross-matched sample of
    $\sim$~$121,000$ objects between J-PAS and DESI.
    The model classifies stars, galaxies, and QSOs and
    estimates photometric redshifts for galaxies and QSOs.
    We demonstrate robust estimation of physical properties for
    galaxies, recovering stellar masses and star formation rates.
    Furthermore, the model accurately predicts equivalent widths
    for major optical emission lines, allowing for the derivation
    of extinction-corrected H$\alpha$ luminosities.
    \texttt{OJALA} successfully reproduces the BPT and WHAN
    diagnostic diagrams, classifying Star-Forming, AGN, and passive
    galaxies.
    The model also provides estimates of stellar atmospheric
    parameters, including effective temperature and metallicity.
    Comparisons with dedicated multilayer perceptrons show that
    a single \texttt{OJALA} model generally matches or outperforms
    specialized predictors, while remaining robust under incomplete
    photometric coverage.
    We release the code, model weights, and a comprehensive
    Value Added Catalog for the J-PAS Early Data Release.
    }
    \maketitle
%

\section{Introduction}
\par Astronomy has unequivocally entered the era of Big Data, characterized by rapidly growing astronomical datasets in both volume and complexity. Large-scale observational projects, such as the Dark Energy Spectroscopic Instrument \citep[DESI;][]{2013arXiv1308.0847L}, the Legacy Survey of Space and Time \citep[LSST;][]{2019ApJ...873..111I}, and the \textit{Euclid} mission \citep{2024arXiv240513491E}, exemplify this transformative shift. Consequently, traditional analytical methodologies face increasing challenges, prompting astronomers to extensively adopt artificial intelligence (AI), particularly machine learning (ML) techniques, for efficient data exploitation.

\par Supervised ML methods have dominated astronomical analyses, effectively predicting photometric redshifts \citep{2019A&A...621A..26P,2021A&A...651A..55S}, estimating physical properties of galaxies \citep{2024MNRAS.531.2011Z,2024MNRAS.530.4260W,2026A&A...705A.219D} and stars \citep{2019A&A...622A.182W,2022A&A...657A..35G}, and classifying astronomical objects \citep[e.g.,][]{2018MNRAS.476.3661D,2021A&A...645A..87B,2025arXiv251120524J}. More recently, the increasing scale and heterogeneity of astronomical datasets have motivated a transition from isolated, task-specific pipelines towards more general and versatile architectures. Such approaches aim to share representations across heterogeneous observations, accommodate incomplete inputs, and address multiple inference tasks within a common framework. Unsupervised and semi-supervised methods provide important examples of this broader trend, demonstrating how representations learned from large datasets can support a range of astronomical applications \citep{2024arXiv240402973W,2024MNRAS.531.4990P,2024arXiv240514930S,2024A&A...688A.160M}.

\par Astronomical analyses fundamentally rely on integrating spectroscopic and photometric datasets. Spectroscopy precisely measures object properties, such as redshifts, emission-line intensities, detailed kinematics, and stellar atmospheric parameters, but often faces biases from targeted preselection and observational limitations. Conversely, photometric surveys systematically capture comprehensive morphologies and spatial distributions over large sky areas, though traditional broad-band (BB) photometry offers limited spectral resolution for detailed physical characterization.

\par Addressing these limitations, the Javalambre Physics of the Accelerating Universe Astrophysical Survey \citep[J-PAS;][]{2014arXiv1403.5237B} uniquely utilizes 54 narrow-band (NB) optical filters plus two medium bands, providing quasi-spectroscopic resolution photometry over extensive areas. This enables detailed and unbiased analysis of galaxies, stars, and quasars (QSOs). The miniJPAS survey \citep{2021A&A...653A..31B}, covering approximately $1$~deg$^2$ in the AEGIS field, has demonstrated the effectiveness of NB photometry for diverse astrophysical studies, including stellar population analyses \citep{2021A&A...649A..79G}, galaxy environment effects \citep{2022A&A...666A..84G,2022A&A...666A.160R}, luminosity and mass function evolution \citep{2024A&A...688A.113D}, characterization of emission-line galaxy populations \citep{2022A&A...661A..99M,2024MNRAS.528.3340B}, active galactic nuclei (AGN) studies \citep{2023A&A...672A.137L}, identification of Ly$\alpha$ emitters \citep{2023A&A...680A..14T}, black hole mass estimations \citep{2022A&A...660A..95C}, determination of stellar atmospheric parameters \citep{2023MNRAS.518.2018Y}, and QSO identification \citep{2023MNRAS.520.3494R,2023A&A...673A.103M,2023A&A...678A.144P,2026A&A...705A.232P}.

\par Comprehensive characterization across galaxies, QSOs, and stars demands accurate redshift determinations, emission-line diagnostics, stellar population analyses, and precise classification techniques. Traditional spectral energy distribution (SED) fitting rapidly becomes inefficient with extensive datasets. While deep learning models offer faster inference, they commonly suffer biases when trained solely on synthetic data and lack flexibility in handling incomplete or partially missing data \citep[e.g.,][]{2021A&A...647A.158M}.

\par To address these challenges, we present \texttt{OJALA}, a transformer-based autoregressive model designed to simultaneously classify astronomical objects and infer their physical parameters from J-PAS NB photometry and complementary BB legacy data. Rather than relying on separate task-specific estimators, \texttt{OJALA} represents photometric measurements and physical quantities as tokens within a common architecture. This formulation allows the model to combine heterogeneous inputs, operate with incomplete observations, and perform multiple classification and regression tasks within the domain of photometric SED analysis. Trained on $\sim 20$ million DESI objects, the model provides robust estimates of stellar masses, star formation rates, and emission-line equivalent widths (EW) for galaxies; photometric redshifts for both galaxies and QSOs; and atmospheric parameters ($T_{\text{eff}}$, $\log g$, [Fe/H], and [$\alpha$/Fe]) for stars. We incorporate unsupervised domain adaptation (UDA) into the training process to align the latent feature distributions of synthetic mocks with real, unlabeled J-PAS observations. This approach is intended to reduce the impact of the sim-to-real gap by encouraging alignment between the latent representations of synthetic and observed photometry. The model's performance is evaluated using a cross-matched sample of J-PAS observations and DESI spectroscopy, which provides the reference classifications and physical parameters used in the evaluation.

\par Throughout this paper, we assume a flat $\Lambda$CDM cosmology with $H_0 = 70\,\mathrm{km\,s^{-1}\,Mpc^{-1}}$, $\Omega_{\mathrm{m}} = 0.3$, and $\Omega_{\Lambda} = 0.7$, and we adopt a \citep{2003PASP..115..763C} IMF for all stellar masses and star formation rates (SFRs) unless stated otherwise. We adopt the AB system for magnitudes.

\par The paper is organized as follows. Sect.~\ref{sec:data} describes the DESI and J-PAS datasets, along with the generation of the synthetic training mocks. Sect.~\ref{sec:model} details the architecture of \texttt{OJALA} and the training strategy. In Sect.~\ref{sec:results}, we present the results for spectral classification, photometric redshift estimation, and the inference of physical properties for galaxies and stars, followed by comparisons with baseline methods. Sect.~\ref{sec:discussion} discusses the broader implications and limitations of the approach. Finally, Sect.~\ref{sec:conclusions} provides a summary and conclusions.

\section{Data and Mocks}\label{sec:data}

\subsection{DESI Spectroscopic Catalogs}
\par Our model is trained on data from DESI \texttt{DR1} \citep{2025arXiv250314745D}. This release includes spectra for over 20 million unique targets, comprising Main Survey observations from May 2021 to June 2022 and reprocessed Survey Validation data. The dataset covers a spectral range of 360--982.4 nm with a resolution of $R\sim2000$ at 360 nm to $R\sim5500$ at 980 nm, providing spectra for approximately 14.1 million galaxies, 1.6 million quasars, and 4.5 million stars. In DESI, \texttt{SPECTYPE} is assigned by the Redrock pipeline and corresponds to the spectral class of the best-fitting template among the main families (\texttt{STAR}, \texttt{GALAXY}, \texttt{QSO}), obtained through template fitting and $\chi^2$ minimization over redshift  (S. Bailey in prep.).

\par We utilize several Value added catalogues (VACs) derived from DESI \texttt{DR1} \footnote{ The detailed description, data access and references for all VACs are provided in the official DESI documentation webpage \url{https://data.desi.lbl.gov/doc/releases/dr1/}.}. Specifically, we employ the \texttt{FastSpecFit} Spectral Synthesis and Emission-Line Catalog, which provides stellar continuum and emission-line modeling for the DESI sample. Our model includes the EW of [O\textsc{iii}] $\lambda$5007, H$\beta$, H$\alpha$, and [N\textsc{ii}] $\lambda$6584, along with the stellar continuum fluxes at H$\alpha$ and H$\beta$ and spectroscopic redshifts. We restrict our input features to this specific set of lines as they constitute the standard diagnostics required for ionization classification via BPT and WHAN diagrams, which form the core of our emission-line analysis. While \texttt{FastSpecFit} also provides measurements for quasars, we restrict this version of the model to the galaxy sample. QSO spectra involve significantly higher complexity due to the presence of broad-line components and non-thermal continua; therefore, their line measurements are left for future work.
\par We incorporate the stellar masses and SFR from the AGN Host Galaxies Physical Properties VAC. This catalog utilizes SED modeling from the \texttt{CIGALE} code, accounting explicitly for AGN contributions \citep{2024A&A...691A.308S}. 

\par For stellar objects, we utilize the DESI \texttt{DR1} Stellar Catalog, which provides radial velocities, elemental abundances, and stellar atmospheric parameters derived via \texttt{RVSpecFit} \citep{2025arXiv250514787K}. Our model incorporates the following parameters: effective temperature ($T_{\text{eff}}$), surface gravity ($\log g$), metallicity ([Fe/H]), and alpha-element enhancement ([$\alpha$/Fe]). To ensure consistency and mitigate systematic effects in the metallicity measurements, we applied the [Fe/H] calibration formula recommended by the authors in the DR1 documentation.
 \par Additionally, we include $r$, $g$, and $z$ photometric magnitudes from the DESI Legacy Imaging Survey together with W1 and W2 photometric bands from the WISE survey \citep{2010AJ....140.1868W}. Finally, we also incorporate the Point Spread Function (PSF) model fitted by \texttt{PSFEX}, which characterizes the morphological shape of each DESI Legacy Survey image \citep{2011ASPC..442..435B}. Sources can thus be classified according to the PSF model as \texttt{DEV}, \texttt{EXP}, \texttt{GGAL}, \texttt{PSF}, \texttt{REX}, or \texttt{SER} upon convergence \footnote{Morphological models follow the parametric light-profile fits used in the DESI Legacy Imaging Survey. These include point sources (\texttt{PSF}), round exponential profiles (\texttt{REX}), exponential disks (\texttt{EXP}), de Vaucouleurs profiles (\texttt{DEV}), and S\'ersic profiles (\texttt{SER}). The \texttt{GGAL} label is not a standard parametric model but rather a catalog-level classification used in some DESI pipelines to denote extended sources that are not well described by a specific parametric profile.}.
\par Due to the J-PAS spectral resolution, we do not include in the model parameters tracing kinematic processes such as stellar velocity dispersion or gas kinematics. The physical parameters selected for the present model have been chosen specifically because they can potentially be inferred from J-PAS and BB photometry. However, this list is not exhaustive, and future versions may incorporate additional parameters. The complexity of the model scales with the number of variables included; thus, in this initial version, we maintain a balanced approach, ensuring the model remains practically useful while capturing the most relevant physical parameters describing QSOs, galaxies, and stars.

\subsection{J-PAS data}\label{subsec:J-PAS_data}

\par In this work, we make use of the second Internal Data Release of the J-PAS, denoted as \texttt{IDR202406} (H. Vázquez Ramió et al. in prep.). The data were acquired between May 2023 and May 2024 using the Javalambre Panoramic Camera (JPCam) mounted on the 2.5m Javalambre Survey Telescope (JST250) at the Observatorio Astrofísico de Javalambre (OAJ) and managed by CEFCA in Teruel, Spain. 
\par The J-PAS photometric system comprises 56 filters designed to provide low-resolution spectroscopy ($R \sim 60$) for every pixel in the field of view. This set includes 54 NB filters, spaced by approximately $100\,\AA$, covering the optical range from $3780\,\AA$ to $9100\,\AA$, with a typical full width at half maximum (FWHM) of $\sim 145\,\AA$. These are complemented by two medium-band filters located at the blue ($u_{\text{JAVA}}$, $\lambda_c = 3497\,\AA$) and red ($J1007$, $\lambda_c = 9316\,\AA$) edges of the system. We often refer to the resulting 56-band photometric sampling as the J-spectrum.
\par Observations are performed using four filter trays (T1--T4), each containing 14 NB filters, while a fifth tray (T5) includes 14 copies of the Sloan $i$ band ($i_{\text{SDSS}}$), used as the reference detection image. Although the total area observed in the $i$ band for this release is $\sim 380\,\text{deg}^2$, the common area covered by all NB filters is $\sim 29\,\text{deg}^2$. The typical $5\sigma$ depth of the NB observations ranges from $\sim22.5$ mag in the blue filters to $\sim22.0$ mag in the reddest ones. The detection band reaches a depth of $i_{\text{SDSS}}\sim24.0$ mag for point-like sources.
\par We utilize the photometric catalogues generated in dual-mode, where source detection is performed on the deep $i_{\text{SDSS}}$ coadded reference image. Forced photometry is then extracted from the remaining bands at the positions of these detections. To ensure high-quality SEDs, we specifically employ the 3-arcsec aperture-corrected photometry (\texttt{APER\_COR\_3\_0}). This metric corrects for aperture losses due to Point Spread Function (PSF) variations across the different filters and has been shown to provide the most accurate color indices for galaxies, and photo-z estimates (Vázquez Ramió et al. in prep.). 
For the training and validation of the \texttt{OJALA} model, we select sources from the  \texttt{IDR202406} that have been observed in at least 50 bands. This selection criterion yields a final sample of 1,436,437 unique objects. This dataset provides the real observational sample required for the domain adaptation of the model and for evaluating its performance against the J-PAS--DESI cross-match (see Sect.~\ref{subsec:crossmatch}). We note that a subset of this data, covering $\sim 17\,\text{deg}^2$ in two patches of the sky around the AEGIS field and the HectoMAP field has been released as the J-PAS Early Data Release (\texttt{EDR}). We are releasing the VACs derived from this work for the \texttt{EDR} sample, followed by the full \texttt{IDR202406} catalogue upon the public release of the data.
\subsection{Synthetic J-PAS Photometry, DESI mocks}\label{subsec:simulated J-PAS}
\par We generate synthetic J-PAS photometric fluxes by convolving the DESI spectra with the J-PAS filter system. Due to the different spectral coverage of DESI and J-PAS, the $u_{\text{JAVA}}$ and $J1007$ filters are excluded. Regarding the normalization strategy, the NB fluxes are normalized to the synthetic $i_{\text{SDSS}}$ magnitude derived directly from the spectra. Consequently, the NB input provided to the model encodes strictly the spectral shape, while the absolute luminosity information is supplied separately via observed BB photometry in AB magnitudes ($g, r, z$ from the DESI Legacy Imaging Surveys and $W1, W2$ from WISE). Given that the Legacy Surveys do not observe in the $i$-band, which is the detection band for J-PAS, we compute a calibrated observed $i$-magnitude to serve as the normalization reference. This is achieved by rescaling the synthetic spectroscopic $i$-flux using a scaling factor defined by the ratio between the observed Legacy $r$-band flux and the synthetic spectroscopic $r$-band flux.
\par Finally, photometric errors for the synthetic J-PAS fluxes are assigned using a similarity-search approach. Initially, we train a model exclusively on J-PAS fluxes from the J-PAS data.  Subsequently, for each DESI object, we apply this trained model to identify the closest object in the embedding space within the J-PAS catalog, assigning the corresponding J-PAS errors to the DESI synthetic fluxes and perturbing the fluxes in accordance with these errors. To ensure our training set realistically reproduces the signal-to-noise (S/N) distribution as a function of magnitude observed in J-PAS, the similarity search is performed within 0.1 magnitude bins in the $i_{\text{SDSS}}$ band.
\par
\begin{figure}[htbp]
  \centering

  \includegraphics[width=\columnwidth]{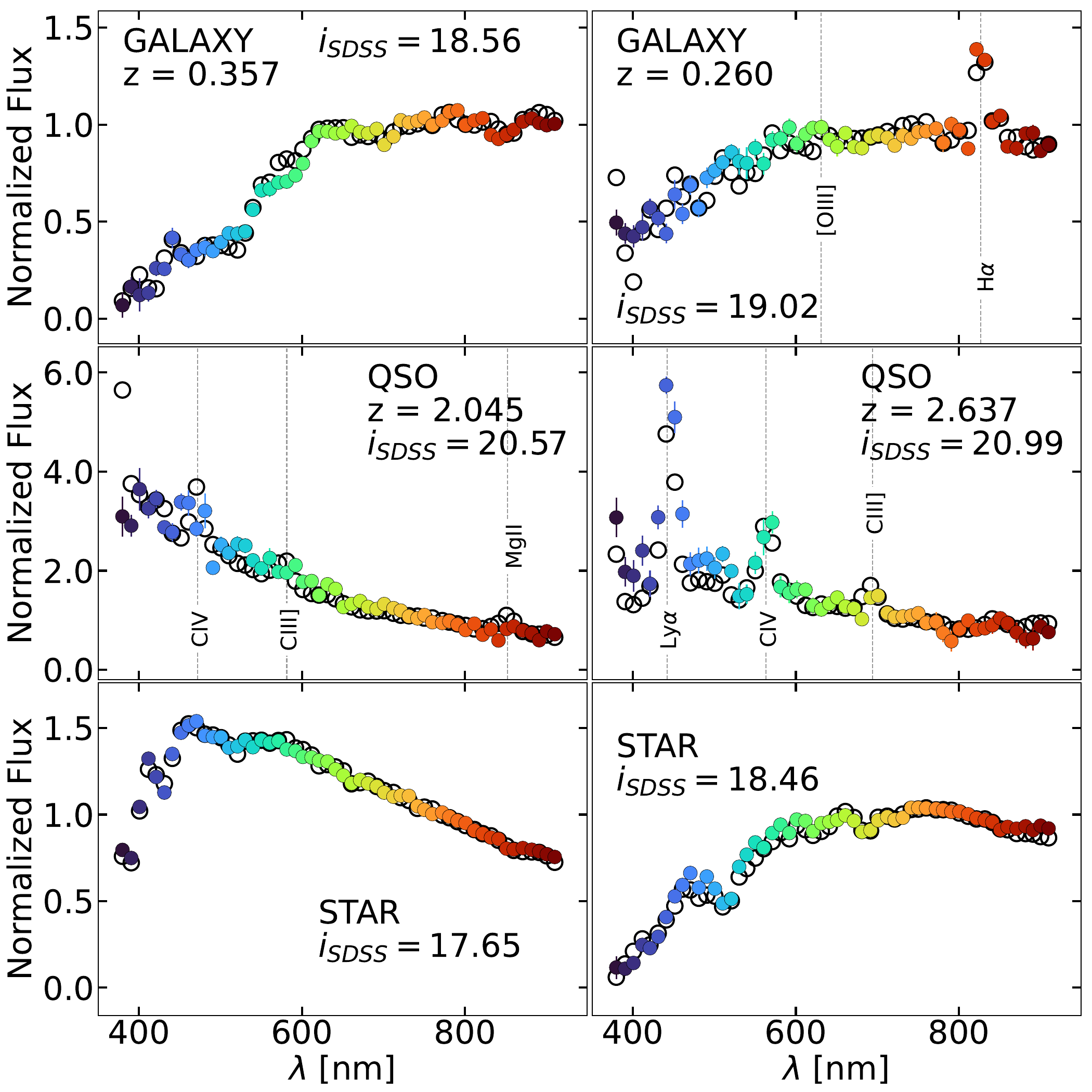}
  \caption{\small{Examples of synthetic J-PAS photometric fluxes from DESI objects (white dots) and their most similar counterparts in the J-PAS dataset (colour dots), identified through a similarity search in the embedding space.}} 
  \label{fig:error_similarity}
\end{figure}

 In Fig.~\ref{fig:error_similarity}, we present illustrative examples of galaxies, QSOs, and stars exhibiting similar features between DESI and J-PAS observations. Note that these matched objects are not cross-matched counterparts as they have been excluded from the training set; rather, they represent different astronomical objects with similar observational characteristics. In contrast to approaches that assign errors solely from survey depth or magnitude–S/N relations, the similarity-based mapping captures better the correlated noise patterns observed in real J-PAS SEDs, which arise from shared observing conditions, calibration effects, and object spectral shape. 

\subsection{J-PAS--DESI cross-match}\label{subsec:crossmatch}
\par To create a robust test sample, we perform a spatial cross-match between the J-PAS \texttt{IDR202406} and the DESI \texttt{DR1} catalogs by matching their coordinates. We exclude objects that do not have at least 50 J-PAS filters observed. This procedure yields a common sample of 121,955 unique objects, which will form our primary test set. This J-PAS--DESI cross-match set is composed of 86,742 galaxies, 26,114 stars, and 9,099 QSOs.

\par From this cross-match, we construct two distinct test sets to evaluate the model's performance under different conditions.
\begin{itemize}
    \item The first set, which we designate \textit{Test-Synth}, is built from the synthetic data. We take the synthetic J-PAS fluxes derived from the DESI spectra for these objects. However, instead of assigning errors via similarity search (as was done for the main training mocks), we assign the exact photometric error from their J-PAS cross-matched counterpart in the \texttt{APER\_COR\_3\_0} aperture. We then perturb the synthetic flux using this specific observational error.
    
    \item The second set, designated \textit{Test-Real}, consists of the actual observed J-PAS photometry for these objects, using the \texttt{APER\_COR\_3\_0} aperture fluxes directly.
\end{itemize}
\par These complementary datasets allow us to directly assess the consistency of the model predictions and quantify the sim-to-real gap by comparing the performance on synthetic versus observed photometry for the same targets. In order to ensure a rigorous and unbiased validation of the model's generalization capabilities, all objects included in this cross-match sample were strictly excluded from the training dataset.

\par
\begin{figure}[htbp]
  \centering
  \includegraphics[width=\columnwidth]{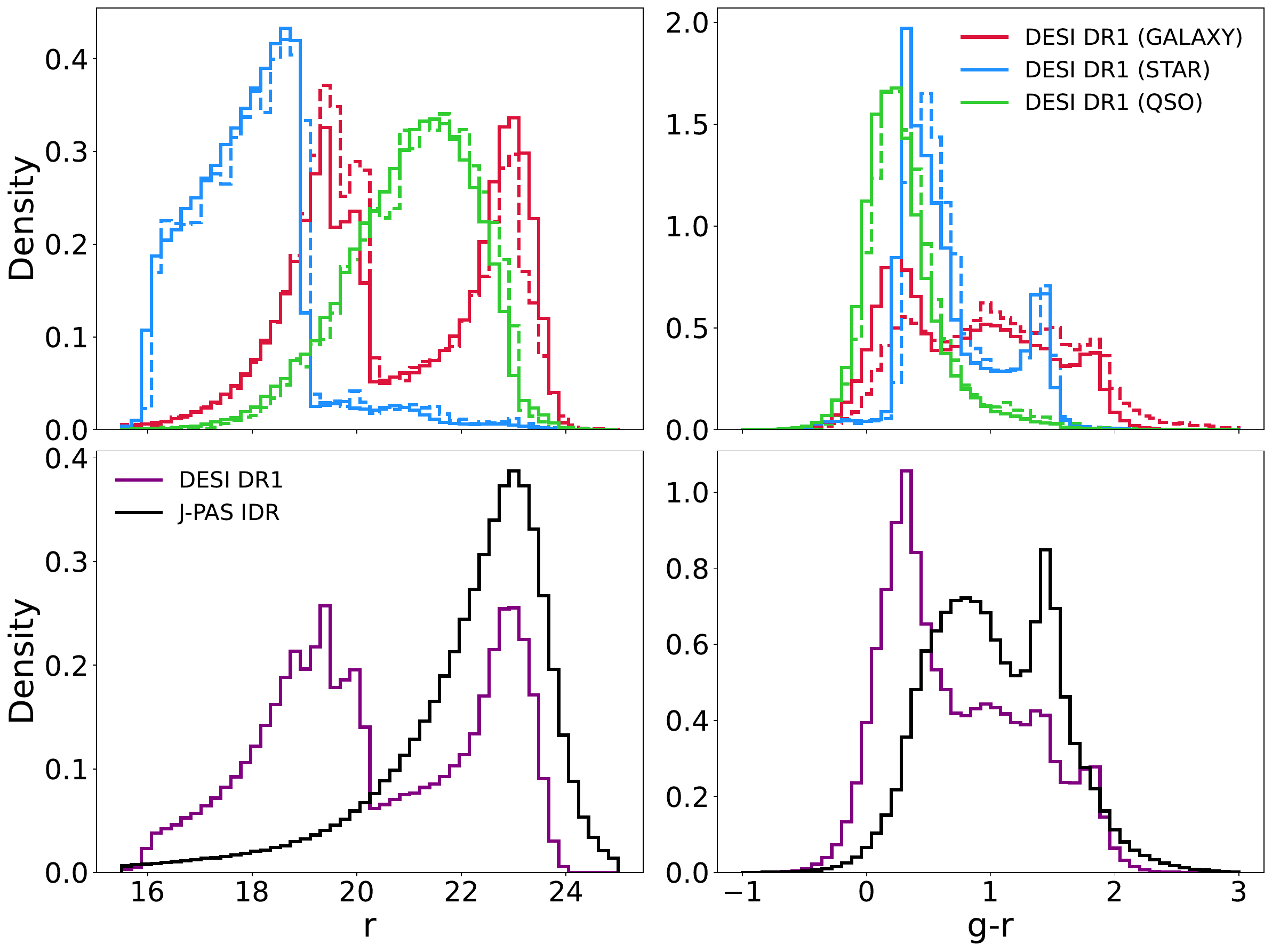}
  \caption{Distribution of the $r$-band magnitude and $g-r$ color obtained from the DESI Legacy Imaging Survey for the datasets used in this work. The top panels show the distributions for the DESI DR1 sample separated into galaxies, stars, and QSOs (solid lines), together with the corresponding populations for the J-PAS--DESI cross-match (dashed lines). The bottom panels compare the overall distributions of the DESI \texttt{DR1} and the J-PAS \texttt{IDR202406} samples.}
  \label{fig:data_distributions}
\end{figure}

 To place the training and observational samples in context, Fig.~\ref{fig:data_distributions} shows the distribution of the $r$-band magnitude and the $g-r$ color obtained from the DESI Legacy Imaging Survey  for the main datasets used in this work. The top panels show the distributions for the three DESI spectral classes separately, alongside the corresponding populations from the J-PAS--DESI cross-match. The bottom panels compare the global distributions of the J-PAS \texttt{IDR202406} sample and the DESI training sample. As expected, the J-PAS--DESI cross-match populations closely follow the distributions of their respective training samples, indicating that the validation set is broadly representative of the domain on which the model was trained. At the same time, the DESI magnitude distribution is shaped by the survey target selection. In particular, stars are mostly limited to bright magnitudes ($r \lesssim 19$), while galaxies display a bimodal distribution reflecting the combination of the Bright Galaxy Survey, which is approximately magnitude-limited to $r \sim 19.5$, together with the Luminous Red Galaxies and Emission Line Galaxies programs that are selected through color cuts. These selection effects should be considered when interpreting the application of a DESI-trained model to J-PAS photometric data. In particular, the spectroscopic training sample may not fully represent the diversity of populations present in the J-PAS catalog, which could introduce biases when extrapolating toward regions of parameter space that are sparsely sampled in the DESI training set. This possibility will require further validation with deeper spectroscopic samples in future work.

\section{The OJALÁ Model}\label{sec:model}

\par Our model builds upon the Transformer-based encoder–decoder architecture introduced by \citet{2024MNRAS.527.1494L}. Originally developed for stellar data, we have enhanced the architecture to handle stars, QSOs, and galaxies, and their associated physical properties within a unified framework. Unlike traditional ML approaches, where the sets of inputs (features) and outputs (targets) are predefined and fixed, \texttt{OJALA} operates analogously to Large Language Models. Instead of predicting missing words based on a textual context, our model predicts missing observations or physical properties of an object given the remaining available observations. This process treats the SED and physical parameters as a flexible sequence of tokens rather than a rigid vector. In one training iteration, the model might predict \ha, stellar mass, and redshift solely from J-PAS NB filters; in another, it might combine BB photometry with stellar parameters to reconstruct the blue filters of J-PAS. This flexibility provides two important advantages. First, the training set can be constructed from the union of all available observations and catalogs, rather than being restricted to a strict cross-match. If a specific observation is missing or a physical property is undefined for an object (e.g., a star lacking a defined stellar mass or a source without a WISE counterpart), those specific tokens are simply excluded from the input-output masking game for that iteration. Second, this methodology enables robust inference on partial data. During the inference phase, the model accepts whatever context is available, allowing it to exploit cross-matched observations when present, while remaining fully capable of performing predictions when they are not.

\par The architecture relies on a Transformer Encoder-Decoder structure powered by the self-attention mechanism. The Encoder ingests the heterogeneous sequence of available photometric measurements, compressing them into a latent representation. The Decoder then acts as a query engine, attending to this latent representation to predict the target values. The core strength of this architecture lies in the attention mechanism, which allows the model to dynamically weight the importance of different inputs based on their physical relevance rather than their fixed position in a list. For instance, when predicting an emission line, the model can pay attention specifically to the NB filters covering the expected wavelength range, effectively learning the spectral correlations required to solve the physical task. We refer the reader to Appendix~\ref{app:architecture} for a detailed mathematical description of the architecture.
\par Since our objective is to simultaneously classify objects and infer their physical properties, \texttt{OJALA} distinguishes between classification and regression tokens. The model is trained by minimizing a composite multi-task loss function. The supervised component, $\mathcal{L}_\text{sup}$, is defined as the sum of three contributions,
\begin{equation}
\mathcal{L}_\text{sup} = \mathcal{L}_\text{reg} + \mathcal{L}_\text{cls} + \mathcal{L}_\text{morph}.
\end{equation}
No explicit weighting is applied between these terms. In practice, the effective contribution of each component is determined by the number of tokens of each type sampled in the batch.
\par For regression tasks, the model is heteroscedastic, meaning it predicts both a mean value ($\hat{y}_i$) and its own intrinsic predictive variance ($\sigma_{\text{pred}, i}^2$). The Negative Log-Likelihood (NLL) loss function is formulated to incorporate both this predicted model uncertainty and the known observational uncertainty ($\sigma_{\text{obs}, i}$) of the training labels. The total variance is modeled as the sum of these two independent components, $\sigma_{\text{total}, i}^2 = \sigma_{\text{pred}, i}^2 + \sigma_{\text{obs}, i}^2$, and the regression loss $\mathcal{L}_\text{reg}$ is defined as the NLL of a Gaussian distribution:
\begin{equation} \label{eq:nll}
\mathcal{L}_\text{reg} = \frac{1}{N_\text{reg}} \sum_{i \in \text{reg}} \frac{1}{2} \left[ \frac{(y_i - \hat{y}_i)^2}{\sigma_{\text{total}, i}^2} + \log(\sigma_{\text{total}, i}^2) \right].
\end{equation}
For classification, standard categorical cross-entropy losses are used for the main object type ($\mathcal{L}_\text{cls}$) and morphological type ($\mathcal{L}_\text{morph}$).

\par A major challenge in training with synthetic data is the sim-to-real domain gap between the synthetic J-spectra (source domain, $S$) and real observations (target domain, $U$) \citep[see, e.g.,][]{2026arXiv260213902L}. To address this, we implement a form of UDA and train the model jointly on J-PAS synthetic data and real, unlabeled J-PAS observations. The J-PAS data are not used as prediction targets; instead, they are introduced only at the encoder level to constrain the latent representation through the domain adaptation objective. In practice, we encourage the latent representations of synthetic and real observations to follow similar distributions, reducing the statistical mismatch between both domains in the encoder feature space. This is achieved by minimizing the distance between the source and target distributions using a weighted Maximum Mean Discrepancy (MMD) penalty, $\mathcal{L}_\text{MMD}$ \citep{10.5555/2188385.2188410}. The details of the kernel functions and the reweighting network used to correct for covariate shift are provided in Appendix~\ref{app:architecture}.

\par The full loss function integrates the supervised and UDA components:
\begin{equation}
\mathcal{L}_\text{total} = \mathcal{L}_\text{sup} + \lambda_\text{MMD} \mathcal{L}_\text{MMD},
\end{equation}
where $\lambda_\text{MMD}$ is a hyperparameter controlling the strength of the domain alignment. Finally, to ensure representative coverage of the full diversity of object types, we assign each training example an individual sampling probability to account for class imbalance and underrepresented subclasses (see Appendix~\ref{app:sampling}).

\section{Results}\label{sec:results}


\subsection{Spectral classification}\label{subsec:spectral_classification}
\par
\begin{figure}[htbp]
    \centering
    \includegraphics[width=\columnwidth]{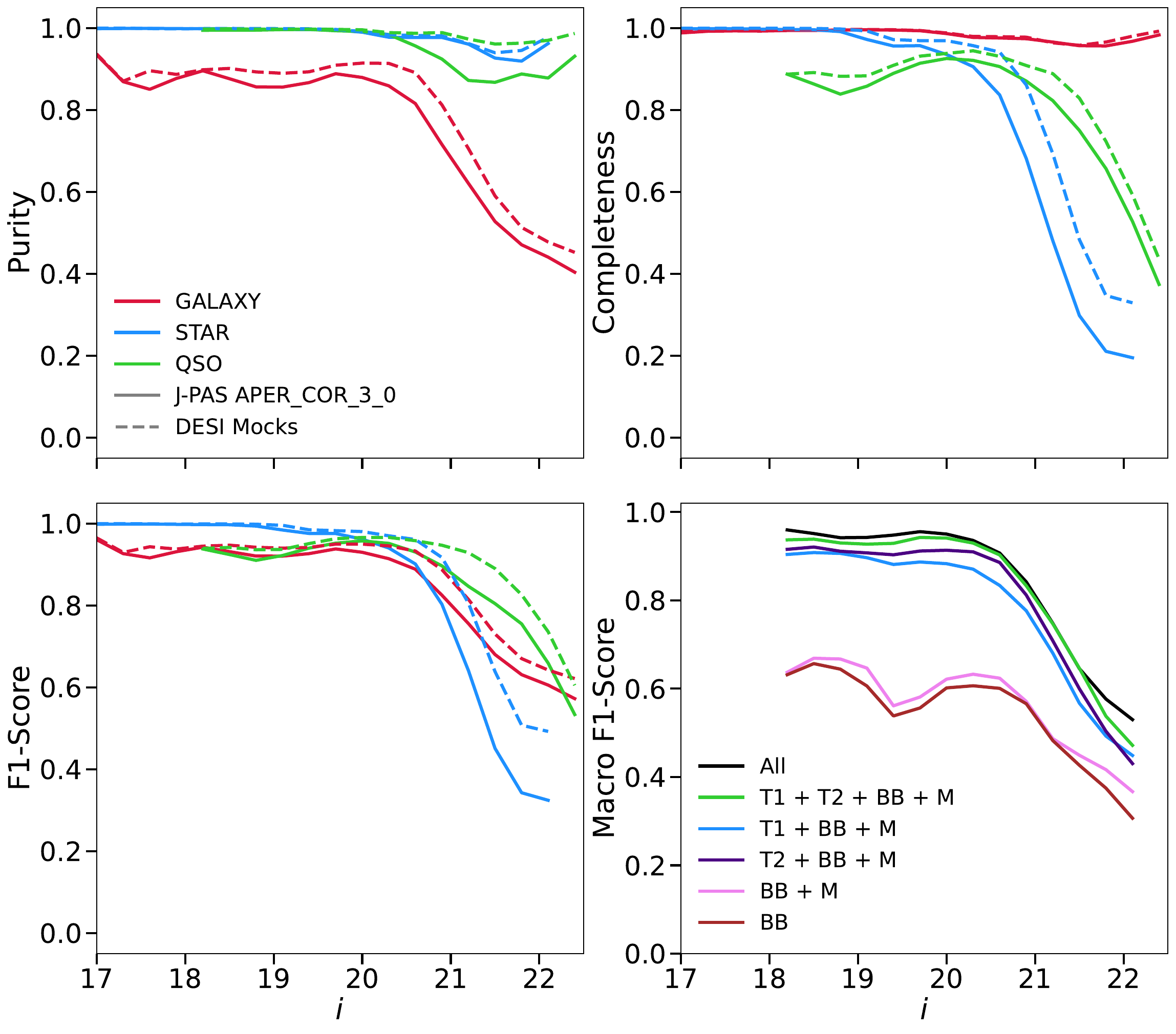}
    \caption{
        Classification performance metrics as a function of the $i$-band magnitude. 
        The top panels and the bottom-left panel compare the results obtained from J-PAS data versus DESI mocks for the three main classes: Galaxies (red), stars (blue), and QSOs (green). 
        We show Purity (top-left), Completeness (top-right), and F1-Score (bottom-left). 
        The bottom-right panel displays the global
        macro-averaged F1-score for different input contexts
        (see Sect.~\ref{subsec:spectral_classification}) applied to the
        real dataset. Metrics are computed in sliding magnitude bins
        centered every 0.3 mag, using a half-width of 0.3 mag.
        Class-wise metrics are shown only when the corresponding
        true class contains at least 100 objects; the macro-averaged
        F1-score is shown only when all three classes satisfy this
        requirement.
    }
    \label{fig:classification_metrics}
\end{figure}

 Figure~\ref{fig:classification_metrics} presents the classification performance of the model as a function of the $i$-band magnitude. To assess the intrinsic model capabilities independent of the class imbalance inherent in the survey, we report balanced metrics, derived from row-normalized confusion matrices where each class is weighted equally. For each magnitude bin, we compute a confusion matrix $C_{ij}$, where rows correspond to the true class $i$ and columns to the predicted class $j$. To remove the effect of the strong class imbalance of the survey, we first row-normalize the matrix,
\begin{equation}
\tilde{C}_{ij} = \frac{C_{ij}}{\sum_j C_{ij}},
\end{equation}
so that each true class contributes equally. We then define the class completeness as
\begin{equation}
\mathrm{Completeness}_i = \tilde{C}_{ii},
\end{equation}
that is, the fraction of objects of class $i$ correctly recovered. The class purity is computed from the same balanced matrix as
\begin{equation}
\mathrm{Purity}_i = \frac{\tilde{C}_{ii}}{\sum_k \tilde{C}_{ki}},
\end{equation}
which measures the fraction of predictions assigned to class $i$ that truly belong to that class after giving equal weight to all true classes. Finally, the class F1-score is defined as the harmonic mean of purity and completeness,
\begin{equation}
\mathrm{F1}_i = \frac{2\,\mathrm{Purity}_i\,\mathrm{Completeness}_i}{\mathrm{Purity}_i + \mathrm{Completeness}_i}.
\end{equation}
For the global metric shown in the bottom-right panel of
Fig.~\ref{fig:classification_metrics}, we compute the
macro-averaged F1-score as the arithmetic mean of the
class-wise F1 values, so that galaxies, stars, and QSOs contribute
equally regardless of their relative abundance in the test sample.
The global metric is reported only in magnitude bins containing
at least 100 objects of each spectral class.
\par As expected, the general trend shows a degradation in performance as objects become fainter and the S/N decreases. However, galaxies exhibit a distinct behavior in Purity compared to stars and QSOs. At the bright end ($i<18$), the purity of galaxies hovers around 0.8; this reflects the relative confusion with low-redshift QSOs, where approximately $20\%$ of bright QSOs are misclassified as galaxies, a contamination that heavily impacts the metric due to the equal weighting of classes. This confusion is physically motivated: the definition of low-redshift QSOs is often ambiguous in the literature. The distinction frequently depends on the selection criteria (e.g., optical emission lines versus X-ray selection) and the continuum between Seyfert I and Seyfert II galaxies, where the visibility of the broad-line region varies  \citep[see e.g.][]{2025A&A...700A.209S}. In this regime, the host galaxy contributes significantly to the total flux, making these objects a composite of both classes. This behavior is consistent with previous findings; \cite{2023A&A...673A.103M} observed similar confusion matrices using an ANN classifier trained for classifying miniJPAS sources and tested on SDSS \citep{2000AJ....120.1579Y} labels. Similarly,  \cite{2024A&A...691A.221D}  noted an increased confusion between QSOs and galaxies at magnitudes brighter than $r\sim18$, attributing this trend to the presence of active galaxies in their samples that exhibit similar characteristics but are categorized differently depending on the reference spectroscopic survey used. Furthermore, since these DESI labels are derived from a best-template assignment, objects lying near the boundary between host-dominated low-redshift galaxies hosting an AGN and classical QSOs may admit similarly plausible fits within both spectral families. In our analysis, however, \texttt{SPECTYPE} is treated as a deterministic ground-truth label, so this intrinsic ambiguity is not propagated into the classification task. A natural extension for future work would be to account for this uncertainty explicitly, for instance by incorporating information from alternative QSO DESI products .

\par Notably, at the faint end ($i > 21$), the galaxy purity drops more sharply than that of the other classes. This indicates that the \texttt{GALAXY} class acts as the default prediction in the low S/N regime. While faint galaxies are rarely misclassified as point sources, faint stars and QSOs are frequently misclassified as galaxies. In a standard abundance-weighted scenario, the overwhelming number of true galaxies would mask these interlopers. However, the balanced purity metric explicitly exposes this asymmetry, revealing that the model tends to default to the galaxy prediction when the input data is ambiguous.

\par The comparison between the solid lines (\textit{Test-Real}) and dashed lines (\textit{Test-Synth}) in Fig~\ref{fig:classification_metrics} reveals a remarkable agreement. The slight performance drop in real data is minimal. The close agreement between the synthetic and real test sets indicates that the resulting training framework yields a small sim-to-real performance gap, allowing the model to generalize effectively to observed photometry. This agreement reflects the complete training pipeline, including the construction of noise-realistic synthetic data and domain adaptation; the individual contribution of these components is not isolated in the present analysis.

\par The bottom-right panel highlights the flexibility of our model
architecture, which can handle missing data and variable input
contexts without retraining. The macro-averaged F1-score
remains close to 0.9 over most of the range up to $i\sim21$,
after which it decreases rapidly as the photometric S/N
deteriorates. We compared different scenarios of data availability. A key finding is that the performance remains practically immutable when the supplementary filter trays (T3 and T4, which include most of the NB red bands) are excluded\footnote{The specific filter compositions for the primary trays are defined as follows. Tray 1 (T1): J0410, J0450, J0490, J0530, J0540, J0550, J0560, J0570, J0580, J0590, J0600, J0610, J0620. Tray 2 (T2): J0390, J0430, J0470, J0510, J0630, J0640, J0650, J0660, J0670, J0680, J0690, J0700, J0710, J0720.}. Even in the scenario using only T1 + BB + Morphology, where BB refers to broad-band photometry comprising the $g$, $r$, $z$ bands from the DESI Legacy Survey and the $W1$, $W2$ bands from WISE, we achieve a macro-averaged F1-score of approximately
0.8 near $i=21$. This can be valuable for follow-up programs such as WEAVE-QSO \citep{2016sf2a.conf..259P,2024MNRAS.530.2688J} that can benefit from J-PAS source classification with reduced filter coverage (e.g., only T1 and T2). Finally, the inclusion of image morphology (comparing BB vs. BB + M) shows a moderate positive effect on the classification, helping to alleviate degeneracies in the separation of point-sources versus extended sources.

\subsection{Photometric redshift estimation}\label{subsection:photo-z}
\par In this section, we evaluate the accuracy of the redshift estimation for galaxies and QSOs. We utilize the metrics defined in the standard literature to quantify the performance: the prediction bias, the normalized median absolute deviation ($\sigma_{\text{NMAD}}$), and the outlier fraction ($\eta$). The normalized redshift error is defined as $\Delta z = (z_{\text{phot}} - z_{\text{spec}}) / (1 + z_{\text{spec}})$. Consequently, the bias is defined as the median of $\Delta z$. The scatter is quantified via the estimator:
\begin{equation}
    \sigma_{\text{NMAD}} = 1.48 \times \text{median}(|\Delta z - \text{median}(\Delta z)|),
\end{equation}
and the outlier fraction $\eta$ is defined as the percentage of objects satisfying $|\Delta z| > 0.15$.
\par We assess the model performance using both the \textit{Test-Synth} and \textit{Test-Real}. While the input context for \texttt{OJALA} includes BB photometry and morphological parameters, tests in which these inputs are removed show that their contribution to estimate photo-z is secondary. Unlike classification, which benefits significantly from morphological information (e.g., separating extended galaxies from point-like stars), redshift estimation is primarily driven by the NB filters. These filters effectively sample spectral features such as emission lines, the 4000 \AA\ break, and other absorption features essential for constraining the redshift.

\par
\begin{figure}[t!]
   \centering
   \includegraphics[width=\columnwidth]{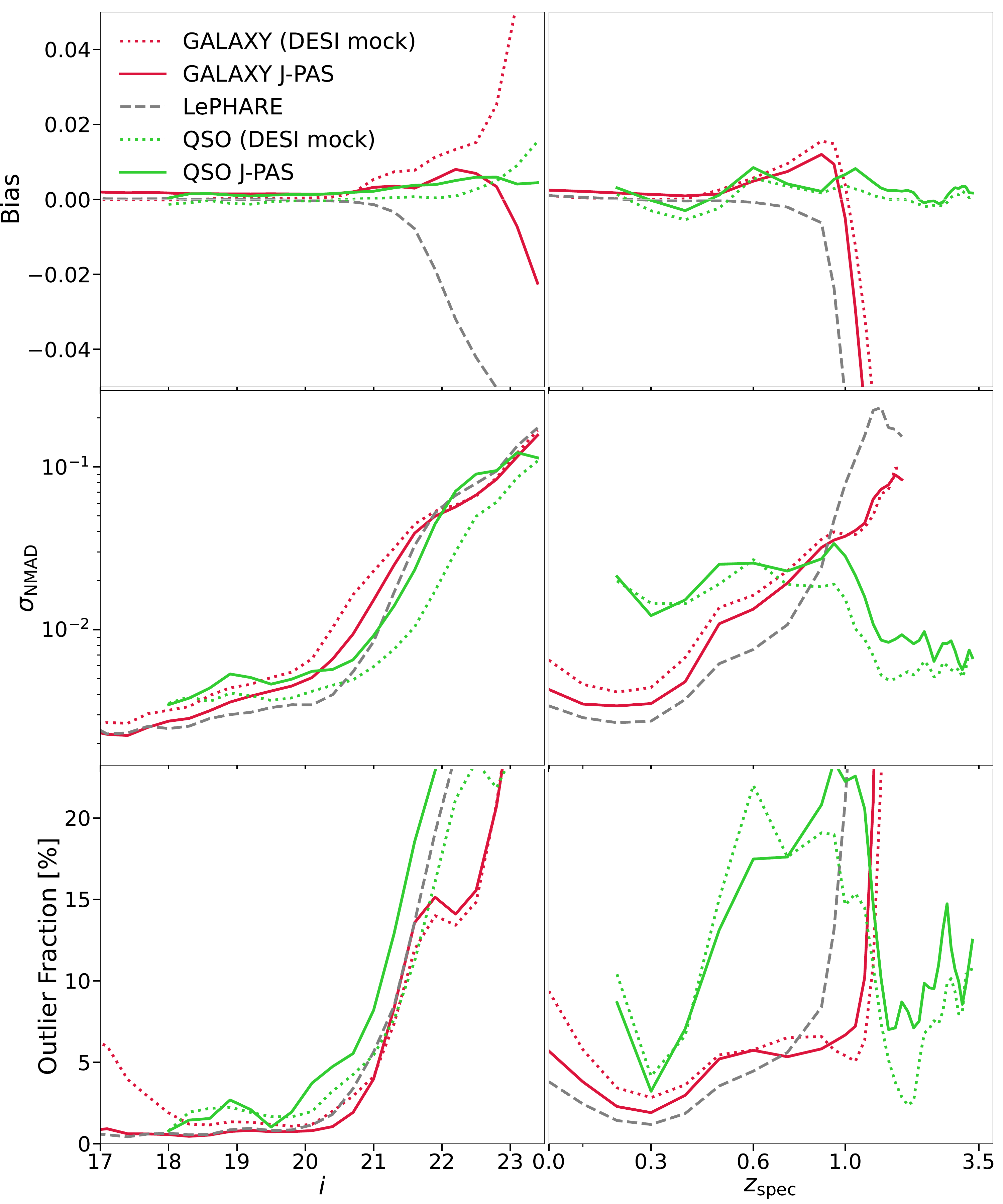}
   \caption{Photometric redshift performance for galaxies (red) and QSOs (green) as a function of $i$-band magnitude (left column) and spectroscopic redshift (right column). The panels show, from top to bottom, the Bias, the normalized median absolute deviation ($\sigma_{\text{NMAD}}$), and the Outlier Fraction (defined as $|\Delta z| > 0.15$). Solid lines represent results on real J-PAS data (\textit{Test-Real}), while dotted lines correspond to the synthetic test set (\textit{Test-Synth}). For galaxies, we also show the performance of the template-fitting code \texttt{LePHARE} (grey dashed lines).}
   \label{fig:photoz_performance}
\end{figure}

 Figure~\ref{fig:photoz_performance} displays the metrics as a function of the observed  $\text{J-PAS}$ $i$-band magnitude (left column) and spectroscopic redshift (right column). Consistent with the classification results discussed in Sect.~\ref{subsec:spectral_classification}, the discrepancy between the \textit{Test-Synth} (dotted lines) and \textit{Test-Real} (solid lines) sets is minor. 
\par As expected, performance degrades with fainter magnitudes due to decreasing S/N ratios. However, for galaxies (red lines), the precision remains high, with $\sigma_{\text{NMAD}} < 0.01$ and an outlier fraction below $5\%$ maintained up to magnitude $i\sim21$. We observe no significant systematic bias up to magnitude $i~\sim~22$. For QSOs (green lines), the scatter is generally larger at bright magnitudes compared to galaxies. This trend is likely associated with low EWs of the emission lines in some sources, which may be insufficient to produce a discernible flux excess in the J-PAS narrow-band filters despite their spectroscopic detection.
\par The redshift dependence (right column) reveals distinct behaviors for the two populations. For galaxies, the highest precision ($\sigma_{\text{NMAD}} < 0.01$) is achieved at $z < 0.6$. While the Balmer break remains within the J-PAS wavelength coverage out to higher redshift, the decreasing S/N of more distant objects progressively limits the achievable precision. The bias remains stable up to $z \sim 0.6$, with a soft increase up to $z \sim 1$, beyond which predictions flatten. Conversely, QSOs exhibit higher scatter and outlier rates at low redshift ($z < 1.0$), likely due to host-galaxy contamination and confusion with stars. However, performance improves significantly at $z > 1.5$, reaching a minimum scatter of $\sigma_{\text{NMAD}} \approx 0.006$ at $z \sim 3.5$. This improvement is physically motivated: at these redshifts, prominent emission lines such as C\textsc{iii}], C\textsc{iv}, and Ly$\alpha$ shift into the optical wavelength range covered by J-PAS, providing strong features for the model to anchor its predictions.

Finally, we compare our results for galaxies with predictions from \texttt{LePHARE}, a standard template-fitting code optimized for J-PAS \citep{2021A&A...654A.101H}. To ensure a fair comparison, we restrict the test set to objects where \texttt{LePHARE} provides a convergent solution, reducing the galaxy sample from 86,742 to 67,257 objects. The comparison (grey dashed lines in Fig.~\ref{fig:photoz_performance}) shows that both methods yield comparable precision. \texttt{LePHARE} achieves marginally lower $\sigma_{\text{NMAD}}$ in intermediate magnitude bins, while \texttt{OJALA} demonstrates greater robustness at the faint end ($i > 22$), exhibiting a lower outlier fraction and a slightly more stable bias.

\subsection{Stellar parameters}\label{subsec:stellar_params}

\par We evaluate the capability of \texttt{OJALA} to infer stellar atmospheric parameters, effective temperature ($T_{\text{eff}}$), surface gravity ($\log g$), metallicity ([Fe/H]), and alpha-element enhancement ([\text{$\alpha$/Fe}]), utilizing the full input context, which includes J-PAS NB filters, BB photometry, and morphological information. The reference values for this analysis are derived from the DESI \texttt{DR1} Stellar Catalog \citep{2025arXiv250514787K}. We restrict our analysis to stars with $i<19$, consistent with the limiting magnitude of the DESI Milky Way Survey program which constitutes the majority of the high-quality spectra in this catalogue. 

\par
\begin{figure}[htbp]
   \centering
   \includegraphics[width=\columnwidth]{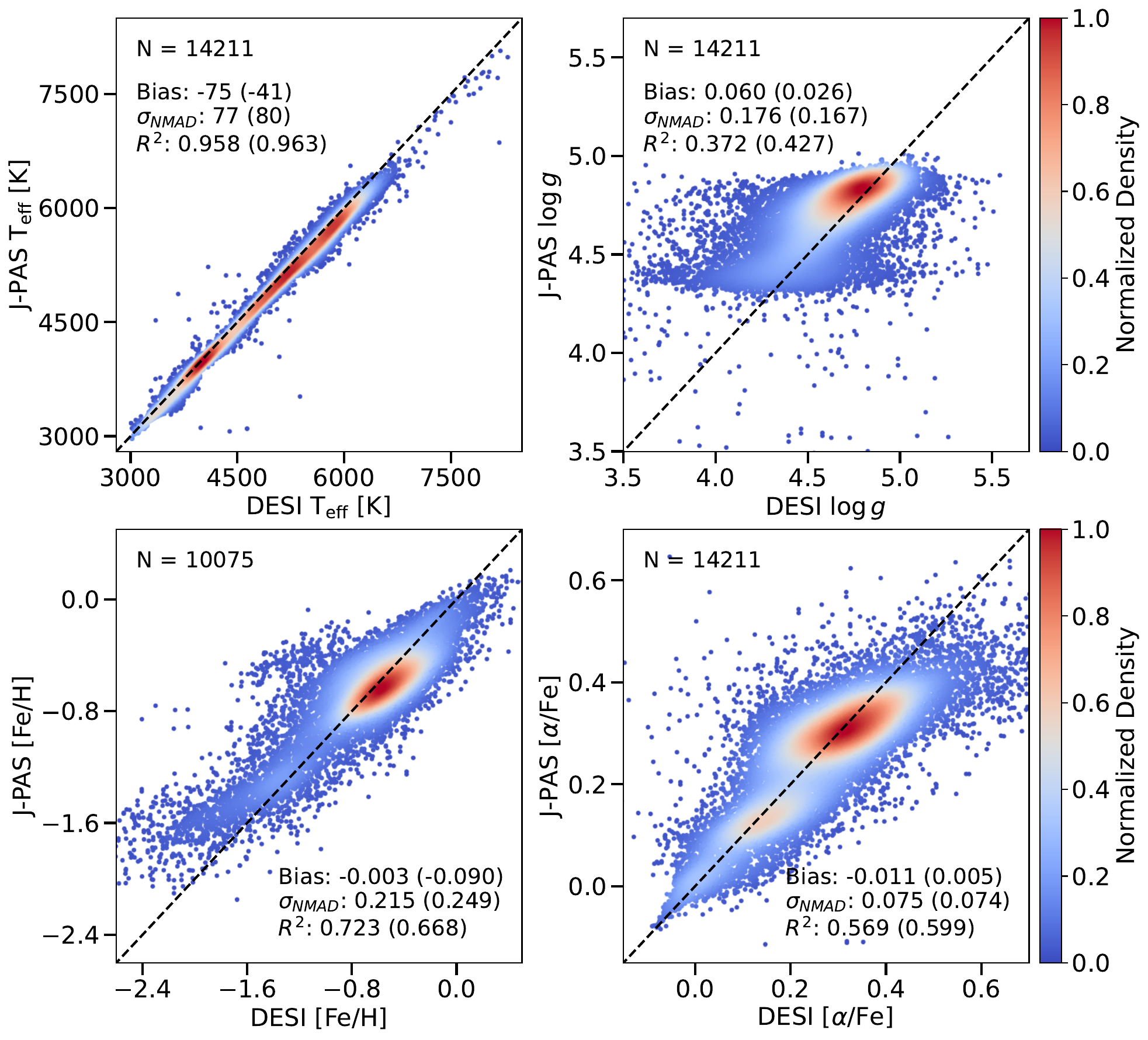}
    \caption{Comparison between DESI spectroscopic parameters and \texttt{OJALA} predictions for stars in the \textit{Test-Real} with $i < 19$. The panels display: Effective Temperature ($T_{\text{eff}}$), Surface Gravity ($\log g$), Metallicity ([Fe/H]), and Alpha-enhancement ([\text{$\alpha$/Fe}]). The points are color-coded by density. The text insets indicate the number of objects ($N$), the Bias, the scatter ($\sigma_{\text{NMAD}}$), and the coefficient of determination ($R^2$) for the real data. Values in parentheses correspond to the performance on \textit{Test-Synth}.}
    \label{fig:stellar_scatter}
\end{figure}

 Figure~\ref{fig:stellar_scatter} presents the one-to-one comparison between the predictions of \texttt{OJALA} and the values derived from the DESI DR1 Stellar Catalog. The metrics reported in each panel correspond to the bias (median difference), the scatter ($\sigma_{\text{NMAD}}$), and the coefficient of determination ($R^2$)\footnote{We adopt the statistical estimators defined in Sect.~\ref{subsection:photo-z}. Unless stated otherwise, for logarithmic variables (e.g., $\log g$, [Fe/H], stellar mass), residuals correspond to the difference in log-space ($\Delta X = X_{\text{pred}} - X_{\text{true}}$), reporting metrics in dex. For $T_{\text{eff}}$, we report absolute residuals in Kelvin. In addition, we report the coefficient of determination, defined as $R^2 = 1 - \sum_i (y_i-\hat{y}_i)^2 / \sum_i (y_i-\bar{y})^2$, where $y_i$ and $\hat{y}_i$ are the reference and predicted values, respectively, and $\bar{y}$ is the mean of the reference values.} for the real J-PAS observations (\textit{Test-Real}), with the corresponding values for the synthetic test set (\textit{Test-Synth}) shown in parentheses.

\par As shown in the top-left panel, $T_{\text{eff}}$ is the most robustly recovered parameter. The model achieves a remarkable precision with minimal scatter. This robustness is expected, as $T_{\text{eff}}$ is the primary driver of the stellar continuum shape (approximated as a black body), a first-order feature that remains detectable even when specific absorption lines are not completely resolved. Nevertheless, the determination of surface gravity (top-right panel) is inherently more challenging. The model performs relatively well for main-sequence stars, which dominate the sample ($\log g \sim 4.0 - 5.0$). However, a significant positive bias ($\sim 0.8$\,dex) is observed for giant stars ($\log g < 3.5$). This suggests a degeneracy where the model struggles to distinguish giants from dwarfs based solely on J-PAS photometry. Physically, robust luminosity class separation requires resolving specific gravity-sensitive features, such as the pressure-broadened wings of Balmer lines or the depth of the Mg\,\textsc{i}\,b triplet and Ca\,\textsc{ii} lines. With a typical FWHM of $\sim 145$\AA, the J-PAS filters essentially dilute these narrow spectral signatures. Consequently, in the absence of high spectral resolution, the information carried by these lines is smoothed out, becoming difficult to disentangle from continuum variations even at high S/N ratios.

\par  For metallicity (bottom-left panel), the model provides accurate estimates with minimal bias in the range $-1.0 < [\text{Fe/H}] < 0.0$, achieving the lowest scatter for solar-like and metal-rich disk stars ($[\text{Fe/H}] > -1.0$). However, it exhibits a slight saturation for metal-poor halo stars ($[\text{Fe/H}] < -1.5$), showing a positive bias. Similarly, alpha-enhancement  (bottom-right panel) is well-constrained for stars with solar-like ratios but shows increased scatter and bias at ($[\text{$\alpha$/Fe}] > 0.4$). Interestingly, the $[\text{Fe/H}]$ for metal-poor stars is recovered with no bias for \textit{Test-Synth}, and we observe a less pronounced deviation of $[\text{$\alpha$/Fe}]$ at high values. Visual inspection of these objects reveals color differences for these stars in the blue part of the spectrum; specifically, DESI spectra tend to be bluer than J-PAS data for these objects. This likely explains the different metallicity estimates from the models, which are tied to the stars' colors.

\subsection{Emission lines in galaxies}\label{subsec:emission_lines}

\par We evaluate the model's capability to predict the EW of the primary optical emission lines: \ha, \hb, \oiii, and \nii, using the cross-matched J-PAS and DESI sample. For this comparison, we restrict the analysis to galaxies with $i < 21$ of which all J-PAS NB filters are observed. As previously demonstrated, this magnitude limit ensures high performance in spectral classification and photometric redshift estimation. Beyond this threshold, the S/N of the J-spectra decreases significantly, causing the photometric contrast for lines with moderate EW to be lost and hindering the model's ability to distinguish emission features from the stellar continuum.

\par
\begin{figure*}
    \centering
    \includegraphics[width=\textwidth]{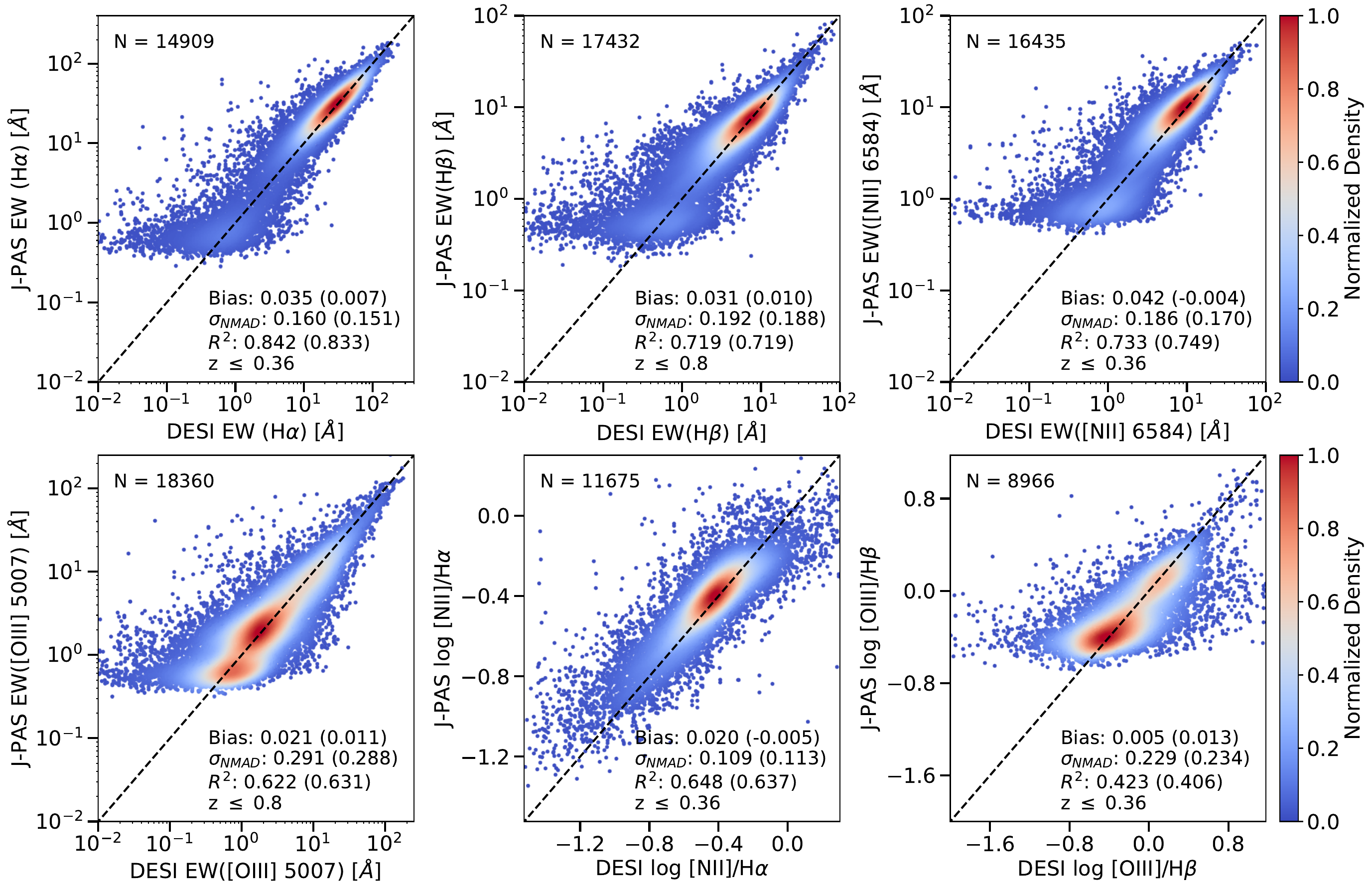}
    \caption{Comparison between DESI spectroscopic EW measurements and \texttt{OJALA} predictions for J-PAS real data (density plots). Metrics for \textit{Test-Real} and \textit{Test-Synth} (in parentheses) are provided in each panel along with the number of galaxies ($N$). The top row displays the hydrogen Balmer lines and \nii, while the bottom row shows \oiii\ and the diagnostic ratios. Statistics include the bias, $\sigma_{\text{NMAD}}$, and $R^2$.}
    \label{fig:EL_scatter}
\end{figure*}

 The performance of the predicted EWs is summarized in Fig.~\ref{fig:EL_scatter}. Specific redshift cuts are applied to ensure that the emission lines are captured by the J-PAS NB filter system. For \ha\ and \nii, we limit the sample to $z < 0.36$ to ensure these features remain within the redmost filter (J0910). For \hb\ and \oiii, the redshift range is extended up to $z < 0.8$.

\par As expected, \ha, being the most intense line in the spectrum, is recovered with the lowest scatter. The performance for \nii\ and \hb\ is comparable to each other. A key advantage of our ML-based approach is its ability to successfully disentangle \ha\ and \nii\ despite their spectral proximity, which frequently causes them to be blended within individual J-PAS filters. We observe a saturation effect around $\sim 1$~\AA, as the model reaches its prediction limit. 
\par The \oiii\ line exhibits the highest dispersion ($\sigma_{\text{NMAD}}$~$\sim$~$0.29$). This is primarily due to its non-Gaussian distribution, characterized by a dominant population of non-emitters and a very long tail extending from $\sim 10$~\AA\ to extreme values. This high dynamic range, reaching even $1000$~\AA\ in Extreme Emission Line Galaxies, presents a significant challenge for ML models as normalization requirements tend to treat the distribution tails as outliers. While our inverse-sampling strategy, based on galaxy densities in the BPT and WHAN diagrams, significantly improves predictions for moderate and high ranges, we observe a flattening of the predictions at high EWs (see Appendix \ref{app:fluxes}). This behaviour reflects a broader limitation of data-driven models: even when sampling strategies attempt to rebalance the training set, objects located in the extreme tails of the parameter space remain statistically rare compared to the bulk population. As a consequence, the model tends to regress toward the more densely populated regions of the distribution, leading to a systematic compression of the most extreme values \citep[see e.g.][]{2022MNRAS.509.4024D}. Consequently, while \texttt{OJALA} can be employed to select Extreme  Emission Line Galaxies, the specific predicted EW values for these extreme outliers should be used with caution (see \ref{fig:flux_residuals}). Other methods based on direct emission-line estimation \citep[e.g.,][]{2022A&A...665A..95I,2025arXiv251121822F,2026A&A...706A.261G} may be more suitable.
\par Emission-line predictions were evaluated within specific redshift ranges to ensure that the features fall within the J-PAS NB filter system. However, to quantify the model's capacity to infer these lines solely from the stellar continuum, we conducted a masking experiment. We re-calculated the predictions by explicitly masking the J-PAS filters corresponding to the expected observed wavelengths of the target lines, thereby deliberately hiding the direct signal from the network. The results, summarized in Table~\ref{tab:masked_stats}, offer a valuable insight: while the model retains reasonable accuracy in the absence of direct photometric evidence, confirming that it learns correlations with the global SED, the performance degrades compared to the standard unmasked configuration. Specifically, the scatter increases across all lines and ratios, accompanied by a slight worsening of the bias. This degradation is also reflected in the coefficient of determination, with $R^2$ decreasing systematically for all emission lines and diagnostic ratios when the corresponding filters are masked. This confirms that \texttt{OJALA} effectively exploits the specific information provided by the J-PAS NB system to refine its predictions, extracting spectral details that are not accessible solely from the continuum.

\begin{table}[h!]
    \centering
    \caption{Comparison of model performance (\textit{Test-Real}) between the standard configuration and the masking experiment. In the masked run, filters containing the emission lines were not provided to the model, forcing it to rely solely on the continuum.}
    \label{tab:masked_stats}
    \resizebox{\columnwidth}{!}{%
    \setlength{\tabcolsep}{5pt}
    \begin{tabular}{lcccccc}
        \toprule
        \multirow{2}{*}{Feature}
        & \multicolumn{3}{c}{Full photometry}
        & \multicolumn{3}{c}{Continuum only} \\
        \cmidrule(lr){2-4} \cmidrule(lr){5-7}
        & Bias & $\sigma_{\mathrm{NMAD}}$ & $R^2$
        & Bias & $\sigma_{\mathrm{NMAD}}$ & $R^2$ \\
        \midrule
        $\log_{10}$ EW(H$\alpha$)
        & 0.035 & 0.160 & 0.842
        & 0.059 & 0.223 & 0.785 \\

        $\log_{10}$ EW(H$\beta$)
        & 0.031 & 0.192 & 0.719
        & 0.040 & 0.235 & 0.676 \\

        $\log_{10}$ EW([NII])
        & 0.042 & 0.186 & 0.733
        & 0.075 & 0.233 & 0.647 \\

        $\log_{10}$ EW([OIII])
        & 0.021 & 0.291 & 0.622
        & 0.026 & 0.346 & 0.526 \\

        $\log_{10}$ [NII]/H$\alpha$
        & 0.020 & 0.109 & 0.648
        & 0.028 & 0.115 & 0.619 \\

        $\log_{10}$ [OIII]/H$\beta$
        & 0.005 & 0.229 & 0.423
        & -0.026 & 0.262 & 0.322 \\
        \bottomrule
    \end{tabular}%
    }
\end{table}

\begin{table}[h!]
    \centering
    \caption{Classification statistics for the WHAN diagram using J-PAS data with different vertical cuts in \nii/\ha. The cuts correspond to values equivalent to \citet{2006MNRAS.371..972S} (S06), \citet{2003MNRAS.346.1055K} (Ka03), and \citet{2001ApJ...556..121K} (Ke01).}
    \label{tab:whan_jpas_stats}
    \resizebox{0.33\textwidth}{!}{%
    \tiny
    \setlength{\tabcolsep}{3pt}
    \begin{tabular}{clccc}
        \toprule
        \mbox{[NII]/H$\alpha$} & Class & Pur (\%) & Comp (\%) & F1 (\%) \\ 
        \midrule
        \multirow{3}{*}{\shortstack{S06 \\ ($\leq$ 0.40)}} 
        & Passive & 93.4 & 88.5 & 90.9 \\
        & SF      & 83.0 & 81.1 & 82.1 \\
        & AGN     & 72.9 & 78.4 & 75.5 \\
        \midrule
        \multirow{3}{*}{\shortstack{Ka03 \\ ($\leq$ 0.48)}} 
        & Passive & 90.7 & 88.5 & 89.6 \\
        & SF      & 76.2 & 89.2 & 82.2 \\
        & AGN     & 76.3 & 65.2 & 70.3 \\
        \midrule
        \multirow{3}{*}{\shortstack{Ke01 \\ ($\leq$ 0.79)}} 
        & Passive & 82.7 & 88.5 & 85.5 \\
        & SF      & 55.0 & 97.9 & 70.4 \\
        & AGN     & 84.6 & 12.7 & 22.1 \\
        \bottomrule
    \end{tabular}%
    }
\end{table}

\par  A direct comparison with our previous work based on Artificial Neural Networks \citep{2021A&A...647A.158M} is shown in Appendix~\ref{app:ANN}. \texttt{OJALA} significantly outperforms this previous, more traditional ML approach, demonstrating the superior capabilities of the transformer architecture and DESI training set.

\par In Fig.~\ref{fig:EL_scatter}, we also show the diagnostic ratios [\text{NII}]/\ha\ and [\text{OIII}]/\hb\ for galaxies where at least one line has an $\text{EW} > 10$~\AA\ and the other has a minimum of $3$~\AA. As expected, the [\text{NII}]/\ha\ ratio is recovered with higher precision than [\text{OIII}]/\hb. We observe a moderate flattening for high values of [\text{NII}]/\ha, a region typically dominated by AGNs.

\par
\begin{figure*}
    \centering
    \includegraphics[width=\textwidth]{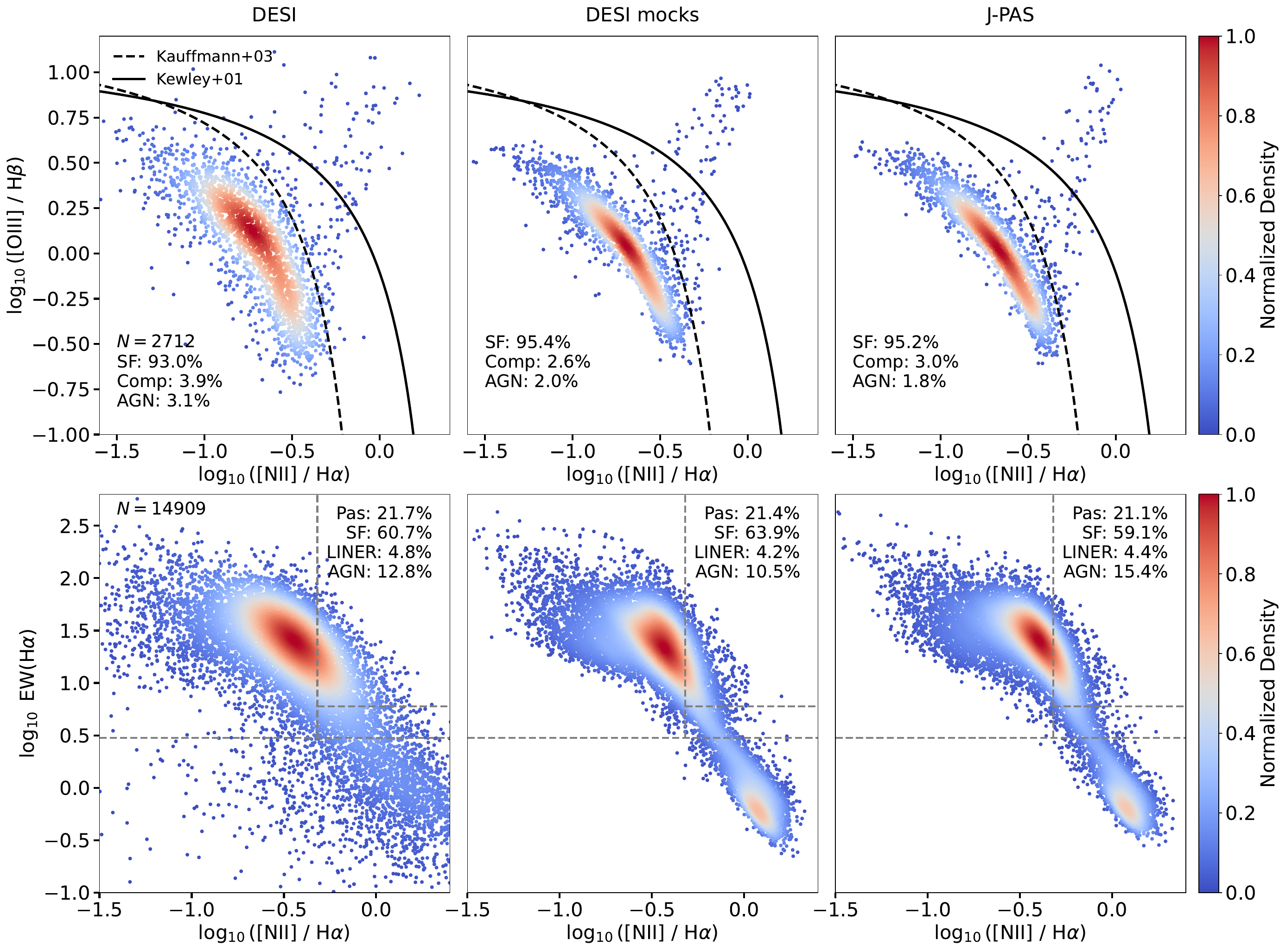}
    \caption{Diagnostic diagrams for galaxy classification. Top row: BPT diagram ($\log_{10}([\text{OIII}]/\text{H}\beta)$ vs. $\log_{10}([\text{NII}]/\text{H}\alpha)$). Bottom row: WHAN diagram ($\log_{10}\mathrm{EW}(\mathrm{H}\alpha)$ vs. $\log_{10}([\text{NII}]/\text{H}\alpha)$). Columns from left to right represent DESI spectroscopic values, \textit{Test-Synth}, and \textit{Test-Real}, colored by point density. The lines indicate the classification boundaries from \citet{2001ApJ...556..121K} and \citet{2003MNRAS.346.1055K}, together with the WHAN transposition of the \citet{2006MNRAS.371..972S} line. Dash-dotted horizontal lines at $\mathrm{EW}(\mathrm{H}\alpha)=6$~\AA\ and $\mathrm{EW}(\mathrm{H}\alpha)=3$~\AA\ define the limits for LINERs and passive galaxies, respectively. Percentages for each population are indicated in each panel, together with the total number of galaxies.}
    \label{fig:BPT_WHAN}
\end{figure*}

 In Fig.~\ref{fig:BPT_WHAN}, we project the predicted line ratios and EWs onto the standard BPT and WHAN diagnostic diagrams. The BPT diagram (top row) is more restrictive, requiring the detection of four lines (\ha, \hb, \oiii, \nii), which limits the validated sample to $\sim~3,000$ galaxies. In contrast, the WHAN diagram (bottom row), requiring only \ha\ and \nii, allows for the classification of $\sim~15,000$ objects. It is important to note that for the WHAN diagram, we do not apply any minimum EW cut on the emission lines. This approach allows us to evaluate the region of passive galaxies (i.e., those without emission lines), a population that cannot be separated in the BPT diagram as they typically fall into the AGN region \citep{2011MNRAS.413.1687C}. Indeed, resolving this degeneracy was the key motivation for the construction of the WHAN diagram.

\par The model accurately reproduces the Star-Forming sequence and identifies intense AGN. However, we note a slight migration of objects toward the bottom-left in the J-PAS panels compared to DESI. This leads to a moderate decrease in the percentage of identified AGN/Composites in favor of the star-forming region, a confusion dominated by the low-emission regime where photometric scatter is higher. It is important to emphasize that while DESI spectroscopic values are used as the reference values, they are representative only of the central 1.5'' fiber, whereas J-PAS uses a 3'' aperture. This difference in physical scale means that the captured gas properties may intrinsically vary between the two datasets.

\par In the WHAN diagram, we observe a high level of agreement in the detection fractions between the synthetic and spectroscopic datasets. However, there is confusion in the low-emission regime, particularly between Low-Ionisation Nuclear Emission-line Regions (LINERs) and passive galaxies. For instance, $\sim 24\%$ of spectroscopic LINERs are classified as passive galaxies, with a similar confusion rate in the opposite direction. Similarly, $\sim 30\%$ of LINERs are classified as AGN. This is largely because the demarcation lines separating LINERs from AGN and from passive galaxies are set at $\text{EW}(\text{H}\alpha)=6$~\AA\ and $3$~\AA, respectively;  in this low-emission regime, the lack of sufficient  photometric contrast makes it intrinsically difficult to distinguish between objects that differ by only a few angstroms in equivalent width. 

\par On the other hand, the model demonstrates high robustness in identifying antagonistic populations, as we do not observe significant confusion of passive galaxies being predicted as star forming (SF) or AGN, nor vice versa. Tables~\ref{tab:bpt_jpas_stats} and \ref{tab:whan_jpas_stats} detail the quantitative results of these classifications. We focus here on the distinction between AGN and SF galaxies (and passive galaxies in the WHAN diagram), explicitly excluding LINERs due to the difficulty of their classification in the low-emission regime. In the BPT diagram, completeness, purity, and F1-scores remain relatively similar regardless of the specific dividing line used, though the best results are obtained with the \citet{2001ApJ...556..121K} (Ke01) line, achieving an F1-score of approximately $80\%$ for SF galaxies and nearly $70\%$ for AGNs (Table \ref{tab:bpt_jpas_stats}). In the WHAN diagram, however, the Ke01 demarcation performs significantly worse than the \citet{2003MNRAS.346.1055K} (Ka03) or \citet{2006MNRAS.371..972S} (S06) definitions due to the saturation of the [\text{NII}]/\ha $ $ ratio at high values. Specifically, AGN completeness drops from $\sim 78\%$ in S06 to just $13\%$ with Ke01 (Table \ref{tab:whan_jpas_stats}). For the S06 and Ka03 lines, we obtain F1-scores of $82\%$ for SF and between $70\%$ and $75\%$ for AGNs, consistent with the BPT results. For passive galaxies, performance remains high with an F1-score around $90\%$.

\par Ultimately, the choice of these separation lines remains a more critical factor for AGN-SF discrimination than the observed percentage differences between datasets. This is particularly evident when comparing WHAN with BPT results; the relatively high percentage of AGN identified in the WHAN diagram ($\sim 24\%$) is partly due to the fact that we have chosen the \citet{2006MNRAS.371..972S} boundary. This limit is notably more conservative than the equivalent Kewley lines used in the BPT space, as it is specifically optimized to ensure a high-purity SF sample, free from any contribution of active nuclear ionization.

\begin{table}[h!]
    \centering
    \caption{Classification statistics for the BPT diagram using J-PAS data. The separation lines correspond to \citet{2001ApJ...556..121K} (Ke01), \citet{2003MNRAS.346.1055K} (Ka03), and \citet{2006MNRAS.371..972S} (S06).}
    \label{tab:bpt_jpas_stats}
    \resizebox{0.3\textwidth}{!}{%
    \tiny
    \setlength{\tabcolsep}{3pt} 
    \begin{tabular}{llccc}
        \toprule
        Line & Class & Pur (\%) & Comp (\%) & F1 (\%) \\
        \midrule
        \multirow{2}{*}{Ke01} 
        & SF  & 67.4 & 99.8 & 80.5 \\
        & AGN & 99.6 & 51.8 & 68.1 \\
        \midrule
        \multirow{2}{*}{Ka03} 
        & SF  & 67.1 & 98.7 & 79.9 \\
        & AGN & 97.6 & 51.6 & 67.5 \\
        \midrule
        \multirow{2}{*}{S06} 
        & SF  & 63.8 & 95.0 & 76.3 \\
        & AGN & 90.2 & 46.1 & 61.0 \\
        \bottomrule
    \end{tabular}%
    }
\end{table}

\par Finally, we also combine the predicted emission-line EWs, continuum flux densities, and photometric redshifts to derive extinction-corrected H$\alpha$ luminosities. The details of this calculation and its validation against DESI spectroscopy are presented in Appendix~\ref{app:fluxes}.

\subsection{Stellar mass and SFR}
\label{subsec:mass_sfr}

\par In this section, we evaluate the model's performance in estimating the stellar mass ($M_{\star}$) and SFR of galaxies based on the SED. We adopt as reference values the estimates from the VAC presented by \citet{2024A&A...691A.308S}. This catalog employs the \texttt{CIGALE} code to model the SED by fitting photometry from the DESI Legacy Imaging Surveys ($g, r, z$) and WISE ($W1, W2$). While the SED fitting relies on BB photometry, it utilizes the precise spectroscopic redshift from DESI, which is essential for breaking the redshift-mass degeneracy and ensuring robust photometric fits. Although this approach does not exploit the full resolution of DESI spectra for stellar population analysis, it ensures a reliable detection of the stellar continuum, particularly for galaxies where the spectroscopic S/N might be insufficient for detailed continuum modeling, and it circumvents the need for aperture corrections. It is important to note that, while we utilize these estimates as reference values for training and validation purposes, they remain model-dependent parameters inferred from BB photometry rather than direct observables.

\par It is important to note how the photometry is handled in our model when deriving extensive physical properties. As detailed in previous sections, the input to the transformer includes the J-PAS \texttt{APER\_COR\_3\_0} fluxes normalized to the $i$-band; this provides the model with the detailed shape of the SED (colors and features). However, to accurately recover extensive quantities such as total stellar mass and total SFR, the model also incorporates the J-PAS $i$-band \texttt{MAG\_AUTO} photometry, along with BB filters from the DESI Legacy Imaging Surveys and WISE bands when available. These magnitudes capture the total light of the object, minimizing aperture effects and allowing the model to properly scale the physical parameters to the total luminosity of the galaxy.

\par
\begin{figure}[htbp]
   \centering
   \includegraphics[width=\columnwidth]{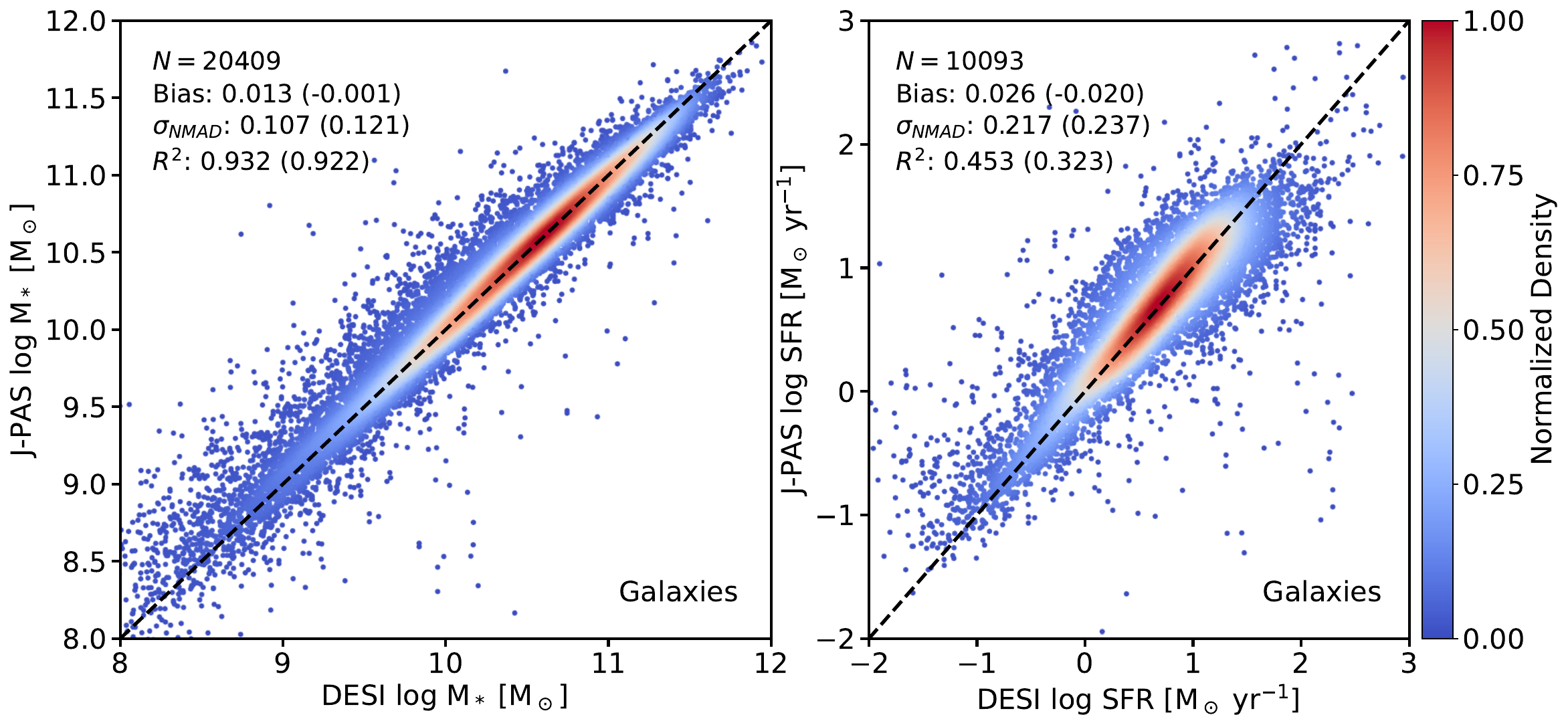}
    \caption{Comparison between the physical parameters derived from \texttt{CIGALE} in DESI and the predictions from \texttt{OJALA} using \textit{Test-Real} for all objects with magnitude below 21. The left panel displays the Stellar Mass ($\log M_{\star}$) and the right panel the SFR ($\log \text{SFR}$) for galaxies. The number of objects ($N$), Bias (median offset), scatter ($\sigma_{\text{NMAD}}$), and $R^2$ are indicated in each panel.}
   \label{fig:logM_logSFR}
\end{figure}

 To assess the accuracy of our predictions, we restrict the analysis to objects with $i < 21$, consistent with previous sections. We do not impose additional cuts on redshift, although the magnitude limit naturally favors the selection of lower-redshift objects. The results are summarized in Fig.~\ref{fig:logM_logSFR}.

\par For galaxies (top row in Fig.\ref{fig:logM_logSFR}), we observe excellent recovery of the stellar mass relative to the reference values, with a bias of $0.013$ dex and a scatter of $\sigma_{\text{NMAD}} \approx 0.11$ dex. Stellar mass is generally a robust parameter in SED fitting, as it is primarily driven by the SED of optical and near-infrared bands in the rest frame. However, accurate mass estimation relies critically on the redshift to determine the correct luminosity distance, information that the \texttt{CIGALE} reference possesses explicitly via $z_{\text{spec}}$. Since our model operates without explicit spectroscopic input, this performance confirms that the J-PAS NB filters provide sufficient constraints on both the redshift and the stellar population properties to match the reference estimates constrained by the spectroscopic redshift.

\par To interpret the specific contribution of the NB photometry beyond the redshift constraint, we performed a control test providing the model with only BB photometry and the exact spectroscopic redshift as inputs. Interestingly, in this scenario, the mass prediction slightly degrades ($\sigma_{\text{NMAD}}$ increases from $0.11$ to $0.12$ dex and the bias shifts to $-0.05$ dex), exhibiting slightly systematic underestimation at both the low ($< 10^9 M_{\odot}$) and high ($> 10^{11} M_{\odot}$) mass ends. This degradation implies that the mapping from BB photometry to stellar mass is not strictly one-to-one; disparate physical configurations (e.g., varying dust and age) can yield identical BB colors, yet result in different mass estimates in the \texttt{CIGALE} analysis. The J-PAS NB system effectively breaks these degeneracies, providing the model with the spectral resolution necessary to distinguish between these cases and correctly recover the specific mass values derived by the reference method, a level of precision unattainable from BB inputs alone even with fixed redshift.

Regarding the SFR (top right panel), we filtered out objects where the uncertainty in the reference estimate exceeds 0.5 dex. This filtering effectively excludes the red cloud population, passive galaxies with specific SFRs below $10^{-11} \text{yr}^{-1}$. In this regime, SED fitting methods often struggle to distinguish between very low star formation rates (e.g., $10^{-2}$ vs $10^{-4} M_{\odot}\text{yr}^{-1}$) because the optical spectrum is dominated by old stellar populations, and the UV flux is negligible. Once these unconstrained objects are removed, the dispersion for Star-Forming galaxies is $\sigma_{\text{NMAD}} \approx 0.22$ dex. The higher uncertainty in SFR compared to mass is expected; SFR is highly sensitive to the age-metallicity degeneracy, dust attenuation laws, and the assumed star formation history (parametric vs. non-parametric), making it a more volatile parameter to retrieve from photometry alone \citep{2013ARA&A..51..393C}. This sensitivity is further evidenced by the BB + $z_{\text{spec}}$ control test, where the SFR retrieval suffers significantly more than the mass: $\sigma_{\text{NMAD}}$ increases from $0.22$ to $0.26$ dex, and the bias rises substantially from $0.026$ to $0.15$ dex. This confirms that SFR retrieval from BB photometry is an ill-posed inverse problem, where the loss of spectral information leads to larger scatter when attempting to reproduce the reference values. The spectral coverage of the J-PAS NB filters, which captures rest-frame UV colors together with emission lines, constrains the spectral shape more tightly, allowing the model to resolve these ambiguities and retrieve SFRs consistent with the reference values derived using the spectroscopic redshift. This interpretation is further supported by the comparison with an independent SED-fitting analysis presented in Appendix~\ref{app:SED_fitting}. There, we show that the SFRs recovered by \texttt{OJALA} are in very good agreement with those inferred using \texttt{BaySeAGal}, indicating that the trends discussed here are not specific to the \texttt{CIGALE}-based reference labels alone.

\subsection{Comparison with supervised MLPs}
\label{sec:mlp_comparison}

We compare \texttt{OJALA} with dedicated supervised multilayer
perceptrons (MLPs), training one independent network per regression
target and a separate three-class classifier. The MLPs use synthetic
training photometry and the corresponding reference labels, with
three hidden layers of 512, 512, and 256 units. Each regressor uses
objects with valid target values and uncertainties and adopts the
same target-specific loss as \texttt{OJALA}; classification uses
cross-entropy. All sources in the real J-PAS test catalogue,
including their synthetic realisations, are excluded from both MLP
training and validation. Normalisation is fitted on the remaining
training data. Validation loss controls learning-rate reduction,
early stopping, and the selection of the best checkpoint across
three training seeds, without using test performance.

To provide a baseline that also learns to handle missing inputs,
the MLPs receive random masking during training. For 75 per cent
of presentations, an object-specific masking probability is drawn
uniformly from $[0,1]$, and each available input is independently
hidden with that probability. The remaining 25 per cent receive
no additional masking. Missing inputs are represented through
explicit availability indicators. Validation uses the available
inputs without additional masking.

Evaluation uses the observed J-PAS--DESI crossmatches and the
target-dependent quality cuts adopted above. We retain 100, 75,
50, or 25 per cent of each object's initially available photometric
bands, including J-PAS, Legacy, and WISE, using identical masks
for both models. Available morphology is retained, and existing
missing values remain missing. No reference physical parameters
or spectroscopic redshifts are supplied as inputs. Metrics at each
incomplete coverage level are averaged over five random masking
realisations, with nested subsets across retention levels.

\begin{figure*}
    \centering
    \includegraphics[width=\textwidth]
    {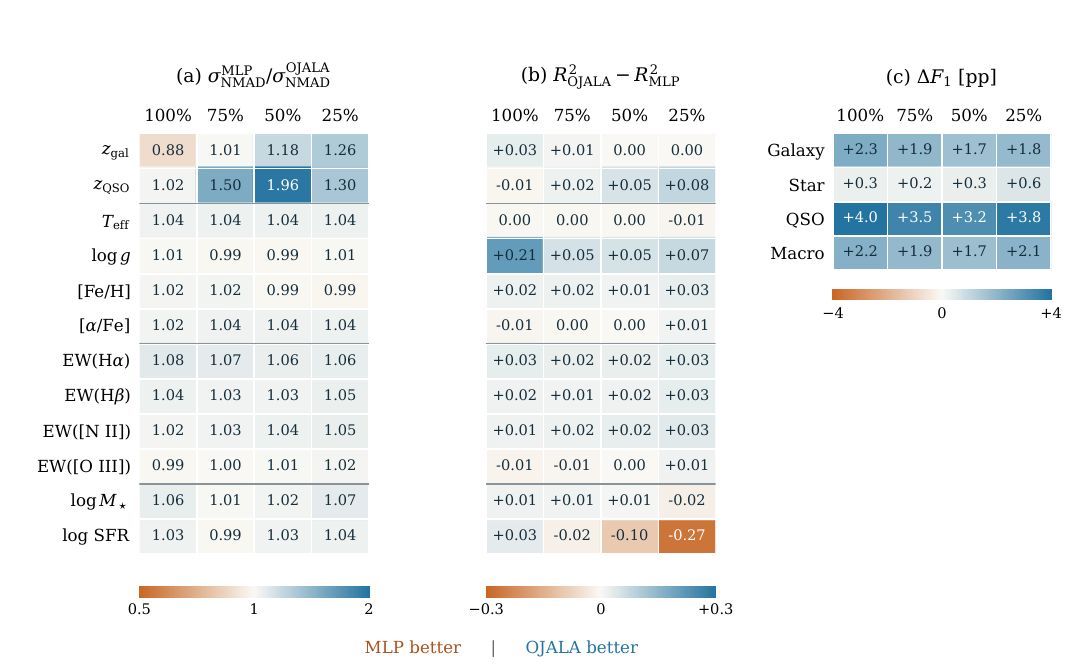}
    \caption{
    Performance of \texttt{OJALA} relative to independent MLPs
    trained with random input masking evaluated on real J-PAS observations.
    Columns indicate the fraction of initially available
    photometry retained; morphology is preserved.
    Panel (a): scatter ratio
    $\sigma_{\rm NMAD}^{\rm MLP}/\sigma_{\rm NMAD}^{\rm OJALA}$.
    Panel (b): $R^2_{\rm OJALA}-R^2_{\rm MLP}$.
    Panel (c): $F_{1,\rm OJALA}-F_{1,\rm MLP}$ in percentage
    points, by class and for the macro-averaged aggregate.
    Class-wise scores are computed from row-normalized confusion
    matrices, following Sect.~\ref{subsec:spectral_classification},
    and the aggregate is their arithmetic mean.
    Blue favours \texttt{OJALA}; orange favours the MLPs.
    Metrics are averaged over five masking realisations,
    except at full retention. Redshift scatter uses residuals
    normalised by $(1+z_{\rm true})$; logarithmic quantities
    are evaluated in dex and $T_{\rm eff}$ in Kelvin.}
    \label{fig:mlp_comparison}
\end{figure*}

Figure~\ref{fig:mlp_comparison} shows that a single \texttt{OJALA}
model generally matches or outperforms the specialised MLPs,
even when these are trained with extensive input masking.
For most physical parameters, the robust scatters are similar,
while \texttt{OJALA} retains an advantage in redshift precision
under reduced photometric coverage. At 25 per cent retention,
the MLP-to-\texttt{OJALA} scatter ratios are 1.26 for galaxies
and 1.30 for QSOs; for QSOs, the ratio reaches 1.96 at
50 per cent retention. Classification consistently favours
\texttt{OJALA} for all three classes and every retention level,
with macro-averaged $F_1$ gains of 1.7--2.2 percentage
points. Thus, its competitive performance with incomplete inputs
persists against baselines explicitly trained for this setting.

The regression metrics also reveal complementary behaviours.
With all available photometry, galaxy redshifts have larger
robust scatter with \texttt{OJALA}, but higher $R^2$.
At 25 per cent retention, $\log {\rm SFR}$ has slightly smaller robust scatter but lower $R^2$, reflecting larger squared errors despite the narrower central residual distribution.
These results demonstrate that a single \texttt{OJALA} model generally matches or outperforms dedicated MLPs, with consistent advantages in classification and improved redshift precision under incomplete photometric coverage. This shows that the unified \texttt{OJALA} framework can retain competitive predictive performance while accommodating heterogeneous and partially observed astronomical data within a single model.

\section{Discussion}\label{sec:discussion}

\par In the era of Big Data in astronomy, one of the most
significant challenges is processing vast volumes of data
efficiently without compromising accuracy. A primary objective
of this work was to demonstrate that a single, unified
Transformer-based model can competently perform diverse
predictive tasks, ranging from object classification to the
regression of multiple physical parameters, that traditionally
require distinct, specialized pipelines. This unification
simplifies the deployment of predictive models for
large-scale catalogue production. \texttt{OJALA} is a relatively
lightweight model ($\sim$4.6 M parameters) compared with
larger astronomical representation-learning models such as
\texttt{AstroPT} \citep{2024arXiv240514930S}, which scale to
hundreds of millions or billions of parameters. While these
models pursue broader representation learning and downstream
adaptation, \texttt{OJALA} is trained directly for classification
and physical-parameter inference within the domain of photometric
SED analysis. This compact architecture enables the massive
production of VACs in very short timeframes; for instance,
processing one million objects takes approximately 45 minutes
on a standard consumer CPU.

\par The comparison with dedicated supervised MLPs
(Sect.~\ref{sec:mlp_comparison}) further shows that this
unification need not come at the expense of predictive
performance. A single \texttt{OJALA} model generally matches
or outperforms the specialized predictors across the evaluated
tasks and remains competitive when the available photometric
coverage is reduced. This is particularly relevant for survey
catalogues with heterogeneous or incomplete photometric
coverage. The comparison should be interpreted at the level
of the complete modelling frameworks, rather than as an
architecture-only ablation, since \texttt{OJALA} also differs
from the MLPs in its shared multi-task formulation and training
strategy.

\par Despite its success, the current implementation has several
important limitations that must be acknowledged. First, for
continuous parameters, the model predicts only
a central value and a scale within a fixed unimodal
parametric distribution. This is not a limitation of
the use of a likelihood-based loss itself, but of the
restricted parametric form used to describe the predictive
posterior. In practice, this means that \texttt{OJALA} provides
a local unimodal approximation, which is adequate for many
well-constrained quantities but cannot fully represent
asymmetric or multimodal posteriors arising from photometric
degeneracies. In addition, autoregressive masked models may
suffer from imperfect calibration and weak correlations between
predicted tokens in high-dimensional spaces
\citep[see e.g.][]{2021arXiv210304922B}. Future versions of
\texttt{OJALA} could address this limitation by adopting a more
expressive generative decoder, such as a diffusion model,
capable of representing more complex predictive
distributions at the cost of a more expensive inference procedure.

\par Another limitation common to deep learning models is the
difficulty in handling rare and out-of-distribution
samples. As observed in Sect.~\ref{subsec:emission_lines},
the model performance degrades at the edges of the parameter
space (e.g., extreme emission-line galaxies). Addressing this
long-tail problem remains an open challenge in data-driven
astronomy.

\par Finally, it is essential to contextualize the validation
results by addressing the intrinsic limitations of the reference
dataset, encompassing both sample selection and parameter
reliability. First, as discussed in
Sect.~\ref{subsec:crossmatch}, the validation relies on the
J-PAS--DESI cross-match, which inherently carries the target
selection biases of the spectroscopic survey. Consequently,
\texttt{OJALA} may exhibit performance biases when applied to
galaxies or populations underrepresented in the DESI training
set, particularly at fainter magnitudes. Future validation using
deeper spectroscopic surveys is necessary to quantify and
mitigate these potential selection biases.

\par Beyond these selection effects, within the sample itself,
we must distinguish between strict ground truth and
model-dependent reference values. While spectroscopic redshifts
and spectral classifications provided by DESI are generally
robust observables, they are not immune to interpretative
ambiguities, as illustrated by the classification challenges
at the interface between low-redshift QSOs and their host
galaxies. Furthermore, the physical parameters adopted as
training labels lie on a spectrum of model dependence.
Emission-line measurements, while derived directly from spectra,
are sensitive to the specific continuum subtraction and fitting
algorithms employed. Stellar atmospheric parameters are
contingent upon the underlying template grids. Most notably,
extensive properties such as stellar mass and star formation
rates are not direct observables but complex inferences heavily
dependent on the assumed priors, star formation histories, and
dust attenuation laws inherent to the SED-fitting codes
(e.g., \texttt{CIGALE}). Consequently, the performance metrics
reported here effectively quantify the model's ability to
reproduce these specific reference values rather than an
absolute physical truth. It is evident that \texttt{OJALA}'s
predictive accuracy is intrinsically bound to the reliability
of these input catalogs; therefore, the future integration of
alternative or improved VACs could improve the quality
of the corresponding predictions.

\par Beyond these limitations, the architecture of
\texttt{OJALA} opens avenues for applications beyond single-object
analysis. A particularly promising direction, which will be
the subject of a forthcoming paper, is the analysis of spatially
resolved galaxies. J-PAS effectively functions as a low-resolution
IFU survey \citep{2025A&A...704A..52R,2026arXiv260108911R}.
The analysis of such data cubes to produce high-level products
such as maps of star formation, stellar mass density, or emission
line distributions requires complex preprocessing, specifically
source detection and segmentation. Standard detection algorithms
based on luminosity profiles often struggle to distinguish
between high-surface-brightness star-forming clumps in spiral
and irregular galaxies and foreground stars. The classification
capabilities of \texttt{OJALA} could be explored to help
resolve these ambiguities at the pixel or region level. Such an
application would require dedicated validation, since spatially
resolved regions differ from the integrated sources used for
training. Furthermore, segmentation remains a critical challenge;
traditional binning techniques
\citep[e.g. Voronoi tessellation;][]{2003MNRAS.342..345C}
aggregate pixels solely to achieve a target signal-to-noise
ratio, often stripping the resulting bins of physical coherence
by merging regions with distinct stellar populations.
The similarity-search approach used in this work to assign
realistic errors to mocks could likewise be explored for
physics-aware segmentation, grouping pixels based on the
similarity of their embeddings rather than just spatial
proximity or luminosity.

\par Although \texttt{OJALA} is optimized for J-PAS observations,
its Transformer-based structure is not intrinsically
restricted to a single photometric survey. The token-based
formulation allows photometry from different surveys and
physical properties from different catalogues to be incorporated
within the same training framework, including cases in which
only subsets of these measurements are available for a given
object. This architectural flexibility should, however,
be distinguished from direct transfer of the released checkpoint
to previously unseen photometric systems. In the current
implementation, each filter has its own learned identity;
therefore, applying the model to a new passband requires
introducing the corresponding filter token, adapting or
retraining the model with data containing that band, and
validating its performance in the new observational domain.

\par Developing representations that can operate across
different instrumental configurations is an active direction
in astronomical machine learning, as illustrated by
OmniSpectra \citep{IslamFox2026OmniSpectra} and universal
spectral tokenization \citep{Shen2025UniversalSpectralTokenization},
which process heterogeneous spectra on their native wavelength
grids. Motivated by the same general problem, we are exploring
future versions of \texttt{OJALA} in which photometric filters
are represented through their transmission curves rather than
through independent learned identities. Such a representation
could facilitate transfer across photometric systems; its
architecture and performance will be presented and validated
in a separate study.

\par A complementary direction is the integration of
multiple observational modalities within a common model.
Recent astronomical foundation-model developments are beginning
to explore this direction
\citep[e.g.,][]{2025arXivSiudek,2025arXiv251017960P},
combining information from photometry, spectra, and images.
A natural extension of \texttt{OJALA} would involve tokenizing
J-PAS image cutouts directly. This could improve
performance in tasks where morphological information is
important, providing a more robust synergy between spectral
energy distributions and spatial features than the current
scalar morphological parameters allow.

\section{Summary and Conclusions}\label{sec:conclusions}

\par In this work, we have presented \texttt{OJALA}, a
transformer-based autoregressive model tailored for
the analysis of J-PAS NB photometry. Using a training set of
$\sim$~$20$ million synthetic J-PAS SEDs derived from DESI spectra
and employing UDA, we have developed a unified framework capable
of simultaneous object classification and physical parameter
estimation. We have validated the model on real J-PAS observations
with $\sim 121,000$ objects,using DESI spectroscopy and DESI-associated VACs for physical parameters.
The main conclusions of this study are summarized as follows:

\begin{itemize}

    \item \textit{General Performance.}
    We demonstrated that a single transformer model can effectively
    address tasks traditionally handled by multiple
    specialized pipelines. \texttt{OJALA} achieves high accuracy
    in spectral classification and performs robust regression
    of redshifts and physical properties, handling missing data
    naturally through its flexible input context.
    
    Comparisons with dedicated MLPs show that a single
    \texttt{OJALA} model generally matches or outperforms specialized
    predictors, while remaining robust under incomplete photometric
    coverage.

    The resulting training framework, which combines noise-realistic J-PAS synthetic data with domain adaptation, yields a small sim-to-real performance gap: model performance on real J-PAS observations remains very similar to that obtained on the corresponding synthetic DESI mocks.

    \item \textit{Spectral classification.} The model achieves a macro-averaged F1-score close to 0.9 over most of the range up to $i\sim21$, with performance decreasing rapidly at fainter magnitudes.
    We verified the flexibility of the architecture, showing that
    high classification performance
    (macro-averaged F1-score $\sim 0.8$ near $i=21$)
    is maintained even when input data is restricted to the
    blue filter trays and BB photometry, together with morphological information. While star-galaxy separation is robust, we identify a physically motivated confusion between low-redshift QSOs and galaxies ($\sim 20\%$) due to the contribution of the host galaxy light.

    \item \textit{Photometric redshifts.} For galaxies with $i < 21$, we achieve a precision of $\sigma_{\text{NMAD}} < 0.01$ with an outlier fraction below $5\%$, performing comparably to a standard template-fitting code (\texttt{LePHARE}). For QSOs, the precision significantly improves at $z > 1.5$ ($\sigma_{\text{NMAD}} \approx 0.006$ at $z \sim 3.5$) as prominent UV emission lines shift into the J-PAS optical window.

    \item \textit{Stellar Parameters.} Effective temperature ($T_{\text{eff}}$) is the most robustly recovered parameter. Metallicity ([Fe/H]) is accurately inferred with minimal bias in the range $-1.0 < [\text{Fe/H}] < 0.0$, while [\text{$\alpha$/Fe}] is well constrained around solar-like ratios. At the edges of the distribution, however, the performance degrades: metal-poor stars show a positive bias in [Fe/H] and highly alpha-enhanced stars exhibit increased scatter. The comparison between \textit{Test-Real} and \textit{Test-Synth} suggests that part of this behavior, particularly for metal-poor halo stars, may be driven by residual  differences in the blue part of the SED rather than by an intrinsic limitation of the model alone. Surface gravity ($\log g$) remains the most difficult stellar parameter to recover, with degeneracies between giants and dwarfs due to the limited spectral resolution of the NB filters to resolve gravity-sensitive features.

    \item \textit{Emission lines and diagnostics.} We report the performance for each line: \ha\ shows a bias of 0.035 and $\sigma_{\text{NMAD}}=0.160$\,dex; \hb\ and \nii\ yield similar results with biases of 0.031 and 0.042, and scatter values of 0.192 and 0.186\,dex, respectively; while \oiii\ displays the largest scatter with a bias of 0.021 and $\sigma_{\text{NMAD}}=0.291$\,dex. Regarding the diagnostic ratios, we retrieve the [\text{N}\,\textsc{ii}]/\ha\ and [\text{O}\,\textsc{iii}]/\hb\ ratios with a bias and scatter of $0.020$ and $0.109$\,dex, and $0.005$ and $0.229$\,dex, respectively. We reproduce the Star-Forming sequence in both BPT and WHAN diagrams. In the BPT space, we achieve F1-scores of $\sim 80\%$ for SF galaxies and $\sim 68\%$ for AGNs. Similarly, in the WHAN diagram, we recover passive galaxies with an F1-score of $\sim 91\%$, while maintaining high classification accuracy for SF ($\sim 82\%$) and AGN ($\sim 76\%$) populations. Finally, we report accurate results of the \ha\ luminosity corrected from nebular extinction, yielding a global scatter of $\sigma_{\text{NMAD}} = 0.289$\,dex and a bias of $0.031$\,dex.

    \item \textit{Stellar Mass and SFR.} We recover stellar masses for galaxies with high precision ($\sigma_{\text{NMAD}} \approx 0.11$ dex) and minimal bias ($0.013$ dex). The SFR is recovered with a scatter of $\sim 0.22$ dex for Star-Forming galaxies. We demonstrate that NB photometry breaks degeneracies present in BB only fitting, matching the precision of methods that use spectroscopic redshifts.

\end{itemize}

\begin{acknowledgements}
G.M.S., J.E.R.M., A.M.C, L.A.D.G., R.M.G.D., R.G.B., and I.M.P. acknowledge financial support from the Severo Ochoa grant CEX2021-001131S funded by MICIU/AEI/\mbox{10.13039/501100011033}. They also acknowledge support from PID2022-141755NB-I00 and from project AST22\_00001\_Subp~26 and 11, funded by the European Union--NextGenerationEU; the Ministerio de Ciencia, Innovación y Universidades; the Plan de Recuperación, Transformación y Resiliencia; the Consejería de Universidad, Investigación e Innovación (Junta de Andalucía); and the Consejo Superior de Investigaciones Científicas. IPR was supported by funding from the grant PID2023-151122NA-I00 by MICIU/AEI/10.13039/501100011033 and by ERDF/EU. HDS acknowledges financial support by RyC2022-030469-I grant, funded by MCIU/AEI/10.13039/501100011033  and FSE+ and the Spanish Ministry of Science. M.P., L.R., AND D.L.C. acknowledge support from French National Research Agency (ANR) under contract ANR-22-CE31-0026. I.M.P. acknowledges support from PID 2022-140871NB-C21. Based on observations made with the JST/T250 telescope at the Observatorio Astrofísico de Javalambre (OAJ), in Teruel, owned, managed, and operated by the Centro de Estudios de Física del Cosmos de Aragón (CEFCA). We acknowledge the OAJ Data Processing and Archiving Unit (UPAD) for reducing and calibrating the data used in this work.LSJ acknowledges the support from CNPq (308994/2021-3)  and FAPESP (2011/51680-6).A.H.-C. acknowledges financial support by grant PID2023-147386NB-I00 funded by MCIU/AEI/10.13039/501100011033 and ERDF/EU.AdP acknowledges support from the MaX CSIC Excellence Award 20245MAX008. I.B. has received funding from the European Union's Horizon 2020 research and innovation programme under the Marie Sklodowska-Curie Grant agreement ID n.º 101059532.

Based on observations made with the JST250 telescope and JPCam camera for the J-PAS project at the Observatorio Astrof\'{\i}sico de Javalambre (OAJ), in Teruel, owned, managed, and operated by the Centro de Estudios de F\'{\i}sica del  Cosmos de Arag\'on (CEFCA). We acknowledge the OAJ Data Processing and Archiving Department (DPAD) for reducing and calibrating the OAJ data used in this work. Funding for the J-PAS Project has been provided by the Governments of Spain and Arag\'on through the Fondo de Inversiones de Teruel; the Aragonese Government through the Research Groups E96, E103, E16\_17R, E16\_20R, and E16\_23R;  grants PID2024-162229NB-I00, PID2024-155572NB-C21, and PID2024-155572NB-C22 (MICIU/AEI/10.13039/501100011033 and ERDF/EU); the Spanish Ministry of Science and Innovation (MCIN/AEI/10.13039/501100011033 y FEDER, Una manera de hacer Europa) with grants PID2021-124918NB-C41, PID2021-124918NB-C42, PID2021-124918NA-C43, and PID2021-124918NB-C44; the Spanish Ministry of Science, Innovation and Universities (MCIU/AEI/FEDER, UE) with grants PGC2018-097585-B-C21 and PGC2018-097585-B-C22; the Spanish Ministry of Economy and Competitiveness (MINECO) under AYA2015-66211-C2-1-P, AYA2015-66211-C2-2, and AYA2012-30789; and European FEDER funding (FCDD10-4E-867, FCDD13-4E-2685). The Brazilian agencies FINEP, FAPESP, FAPERJ and the National Observatory of Brazil have also contributed to this project. Additional funding was provided by the Tartu Observatory and by the J-PAS Chinese Astronomical Consortium.
\end{acknowledgements}
\section{Data Availability}
\par We release a comprehensive VAC for the J-PAS \texttt{EDR}, encompassing classification, redshifts, and physical parameters for galaxies, QSOs, and stars. We also make public the model weights, training code, and the full set of DESI-based J-PAS mock catalogs. These resources are available at \url{https://github.com/gimarso/OJALA/}.

\bibliographystyle{aa}
\bibliography{aa}

\appendix

\section{Architecture and Training Details}\label{app:architecture}

\par \texttt{OJALA} is a Transformer-based model implemented in \texttt{PyTorch} that utilizes \textit{Flash Attention} \citep{2022arXiv220514135D}, an IO-aware exact attention algorithm. This implementation reduces memory complexity from quadratic to linear with respect to sequence length, facilitating the processing of the high-dimensional context of 54 J-PAS NB  combined with BB photometry from DESI and WISE, and the spectroscopic catalogs. \texttt{OJALA} contains 4.6 million parameters and was trained for 1000 epochs on a single NVIDIA A100-PCIE-80GB GPU, with a total training time of approximately 8 days.

\subsection{Model Specification}
\par The architecture follows an Encoder-Decoder structure. The hyperparameters were selected to maximize capacity while maintaining a compact footprint suitable for rapid inference. The embedding dimension is set to $d_{\text{model}} = 128$.
\begin{itemize}
    \item Encoder: Comprises 2 transformer blocks. Each block utilizes 16 attention heads and a Feed-Forward Network (FFN) with a hidden dimension of 1024. The dropout rate is set to 0.1.
    \item Decoder: Comprises 3 transformer blocks. Each block utilizes 16 attention heads and a wider FFN with a hidden dimension of 2048.
\end{itemize}
\par The total number of trainable parameters is approximately
$4.6$ million. We adopted a compact architecture to
limit computational costs while supporting the multiple
prediction tasks considered in this work. This choice was
motivated by practical considerations rather than a systematic
optimization of model size and training budget. A detailed
study of scaling behaviour is left for future work.
The current size allows for the inference of the entire
J-PAS catalog in a matter of hours on consumer-grade hardware.

\subsection{Physics-Aware Embeddings}
\par To explicitly encode the physical correlations between photometric bands within the latent space, we implement Physics-Aware Embeddings. A standard learnable embedding vector represents the token ID. To this, we add a metadata embedding derived from the physical properties of the filter.
\par For each filter, we normalize its central wavelength $\lambda_c$ to the range $[0, 1]$ (logarithmically scaled) and define a boolean flag indicating whether the band is narrow or broad. These metadata are projected through a lightweight multi-layer perceptron and added element-wise to the token embedding. 
\par Furthermore, we replace the standard sinusoidal positional encoding with a learnable \textit{Relative Positional Bias}. This bias is injected directly into the attention mechanism and is defined as a function of the wavelength separation $\Delta\lambda_{ij} = |\lambda_i - \lambda_j|$ between any two tokens. The model learns distinct Gaussian kernels to describe the interaction between filter types (e.g., narrow-narrow, broad-broad), allowing the mechanism to weigh spectral proximity dynamically.

\subsection{Masking and Sampling Strategy}
\par The training objective is an autoregressive masked modeling task. To prevent the model from learning trivial correlations, we impose constraints on the context generation:
\begin{enumerate}
    \item  Mutual Exclusion of Physical Targets: When the model predicts a physical parameter (e.g., Stellar Mass), other physical labels (e.g., SFR) are masked from the input context. The model is forced to infer the property solely from the observational tokens (photometry) and morphological tokens. Conversely, when predicting missing photometry, any combination of other bands and physical parameters is allowed in the context.
    \item  Context Masking: To prevent overfitting and increase robustness against missing data, we apply a randomized masking to the input context. For every training sample, we retain a random fraction of the available input tokens, imposing a minimum context size of 5 tokens. This forces the model to rely on partial information, simulating real-world scenarios where observations may be incomplete.
    \item  Multi-Token Prediction: In each forward pass, the model predicts 5 target tokens simultaneously. This setup allows the architecture to capture correlations between parameters without enforcing explicit physical constraints.
    \item Inverse Probability Token Sampling: The J-PAS dataset is imbalanced at the token level; for instance, properties such as redshift are defined for the vast majority of extragalactic sources (galaxies and QSOs), whereas specific spectral features like H$\alpha$ emission are only present in a subset of the galaxy population. To correct this, we calculate a sampling weight $w_t$ for each token type based on its inverse frequency in the dataset. These weights are used to increase the probability of selecting rare tokens as prediction targets and decrease the probability of them being masked out from the context.
\end{enumerate}

\subsection{Loss Function}
\par \texttt{OJALA} optimizes a multi-objective loss function. For classification tokens, we use standard Cross-Entropy. For regression tokens, we adopt the Heteroscedastic Loss formulation described in Eq.~\ref{eq:nll}.
\par While standard physical parameters minimize the Negative Log-Likelihood of a Gaussian distribution incorporating observational errors, for redshift estimation we employ a specialized variant of this loss. In this case, the residuals are explicitly normalized by $(1+z)$, directly optimizing the metric $\Delta z / (1+z)$ typically used to evaluate photometric redshift performance.

\subsection{Domain Adaptation}\label{app:subsec:domain}

\par In order to mitigate the distribution shift between the labeled source domain $\mathcal{D}_S = \{\mathbf{h}_S^{(i)}\}_{i=1}^{N_S}$ (synthetic DESI mocks) and the unlabeled target domain $\mathcal{D}_T = \{\mathbf{h}_T^{(j)}\}_{j=1}^{N_T}$ (real J-PAS data), we employ a weighted Maximum Mean Discrepancy (MMD) framework. Unlike standard adversarial adaptation, we focus on density ratio estimation to reweight the target samples, ensuring that the weighted target distribution matches the source distribution in the latent feature space.

The reweighting coefficients $\mathbf{w} \in \mathbb{R}^{N_T}$ are estimated via an auxiliary MLP, $\psi$, normalized by a softmax function with temperature $\tau$:
\begin{equation}
    w_j = \frac{\exp(\psi(\mathbf{h}_T^{(j)}) / \tau)}{\sum_{k=1}^{N_T} \exp(\psi(\mathbf{h}_T^{(k)}) / \tau)}.
\end{equation}

The domain adaptation loss is defined as the squared weighted MMD distance:
\begin{equation}
    \mathcal{L}_{\text{MMD}}^2 = \left\| \frac{1}{N_S} \sum_{i=1}^{N_S} \phi(\mathbf{h}_S^{(i)}) - \sum_{j=1}^{N_T} w_j \phi(\mathbf{h}_T^{(j)}) \right\|_{\mathcal{H}}^2,
\end{equation}
which is empirically computed using a Gaussian RBF kernel $k(\mathbf{x}, \mathbf{y}) = \exp\left(-\frac{\|\mathbf{x} - \mathbf{y}\|^2}{2\sigma^2}\right)$. We employ a dynamic bandwidth strategy where $\sigma$ is re-estimated for each batch using the median heuristic on the combined set of pairwise distances, $\sigma = \text{median}(\|\mathbf{h} - \mathbf{h}'\|)$, ensuring gradient flow across varying latent scales.

Expanding the squared norm yields three terms, which we separate for computational stability:
\begin{equation}
\begin{split}
    \mathcal{L}_{\text{MMD}}^2 = & \frac{1}{N_S^2} \sum_{i,i'} k(\mathbf{h}_S^{(i)}, \mathbf{h}_S^{(i')}) \\
    & + \sum_{j,j'} w_j w_{j'} k(\mathbf{h}_T^{(j)}, \mathbf{h}_T^{(j')}) \\
    & - \frac{2}{N_S} \sum_{i,j} w_j k(\mathbf{h}_S^{(i)}, \mathbf{h}_T^{(j)}).
\end{split}
\end{equation}

The source samples $\{\mathbf{h}_S^{(i)}\}$ in each batch are constructed following the probability sampling strategy detailed in Appendix~\ref{app:sampling}. This strategy creates a balanced source distribution where galaxies, stars, and QSOs are represented with equal probability, stripping away the natural class imbalance of the sky. Consequently, by minimizing $\mathcal{L}_{\text{MMD}}$, the reweighting network $\psi$ is driven to assign higher weights $w_j$ to the minority populations in the real J-PAS batch. This procedure aims to align the probability distribution of the real data with the physics-informed, balanced prior of the DESI mocks, helping the model to generalize across object types regardless of their observational abundance.

To prevent weight degeneracy (where the model selects only a few target samples to match the source), we augment the objective with regularization terms. We maximize the entropy of $\mathbf{w}$ to encourage smoothness and enforce a minimum Effective Sample Size (ESS) to avoid mode collapse:
\begin{equation}
    \mathcal{L}_{\text{DA}} = \lambda_{\text{MMD}} \left( \mathcal{L}_{\text{MMD}}^2 + \lambda_{\text{ent}} \mathcal{R}_{\text{entropy}}(\mathbf{w}) + \lambda_{\text{ess}} \mathcal{R}_{\text{ESS}}(\mathbf{w}) \right).
\end{equation}
In our implementation, we set the domain adaptation weight to $\lambda_{\text{MMD}} = 0.2$. The regularization coefficients are set to $\lambda_{\text{ent}} = 0.1$ and $\lambda_{\text{ess}} = 0.05$. The ESS penalty term is activated only if the effective sample size falls below $70\%$ of the batch size.

\subsection{Training and validation history}
\label{app:training_history}

\begin{figure}[t]
    \centering
    \includegraphics[width=\columnwidth]
    {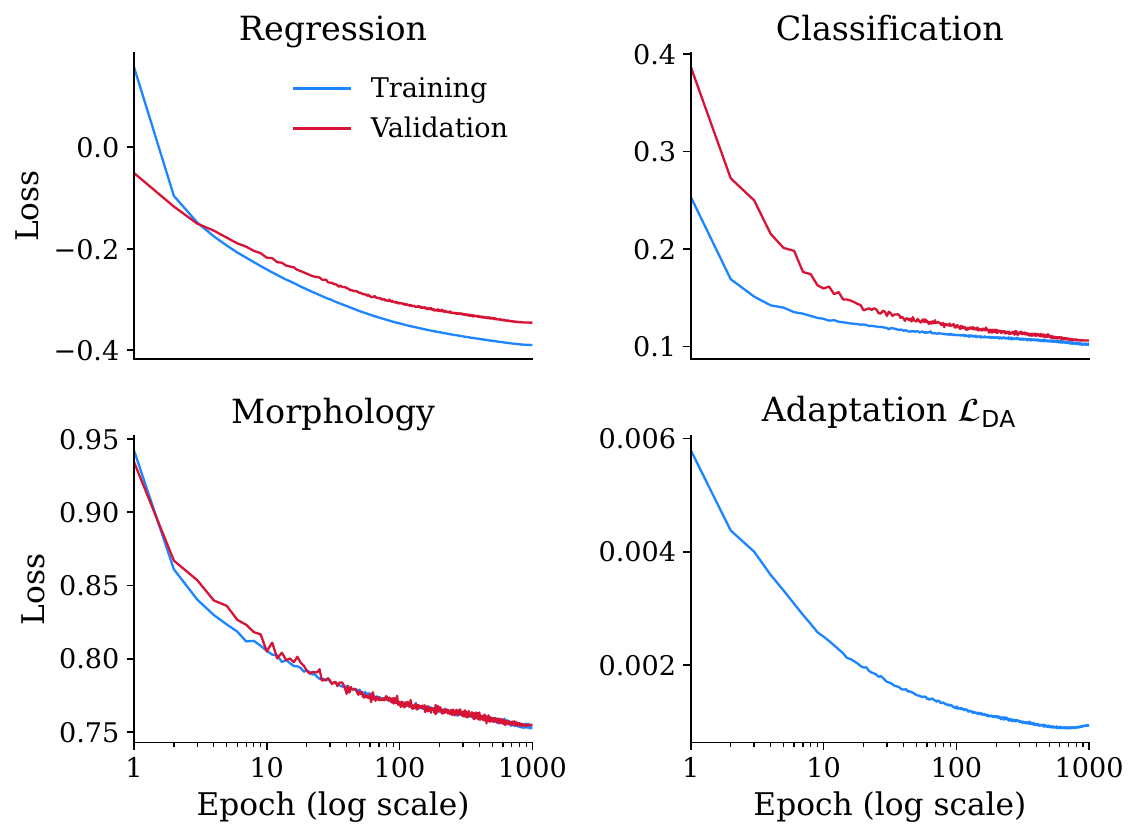}
    \caption{
    Training (blue) and validation (red) losses for the
    \texttt{OJALA} training. Panels show
    regression, spectral classification, morphology and
    the full domain-adaptation contribution
    $\mathcal{L}_{\mathrm{DA}}$, including its regularisation
    terms and overall weight. Adaptation is shown for
    training only. Curves display recorded epoch values
    without smoothing, with a logarithmic epoch axis
    and linear loss axes.
    }
    \label{fig:training_history}
\end{figure}

The validation set consists of 100,000 objects that are kept
completely independent from the training set and are not used
at any stage of model training. During training, repeated presentations of the same astronomical source do not correspond to identical training examples: the available input context and prediction targets are randomly resampled, so a given source can contribute different combinations of input and target tokens across epochs. Figure~\ref{fig:training_history}
shows the regression, spectral-classification, and morphology
losses for both the training and validation sets, together with
the full domain-adaptation contribution, which is evaluated for
the training set only.

The supervised losses decrease rapidly during the initial
epochs and continue to improve more gradually at later stages,
approaching a plateau towards the end of training. The validation
losses closely follow the same overall evolution as the training
losses and remain stable throughout the monitored interval. In
particular, we observe no systematic increase or sustained
deterioration in the validation losses relative to the training
losses, indicating no evidence of overfitting during training.
The run was completed at the prescribed budget of 1000 epochs;
no validation-based early-stopping criterion was applied.

\section{Probability sampling strategy}\label{app:sampling}

\par Deep learning models trained on astronomical datasets often suffer from biases inherent to the training data distribution. The underlying population of observed objects is usually dominated by specific types (e.g., low-mass stars, passive galaxies, or low-redshift objects), leading to a priors-driven performance where the model excels at predicting common properties but fails to generalize to rare or extreme objects. To mitigate this, we implement a sampling strategy based on inverse density weighting. This ensures that the model encounters a balanced representation of the physical parameter space during training.

\par We define the sampling probability, $w_i$, for an object $i$ as inversely proportional to the local density of objects, $\rho(\mathbf{x}_i)$, in a given feature space $\mathbf{x}$:
\begin{equation}
    w_i \propto \frac{1}{\rho(\mathbf{x}_i) + \epsilon},
\end{equation}
where $\epsilon$ is a smoothing term to prevent numerical divergence in empty regions. The local density $\rho$ is estimated using a Gaussian Kernel Density to optimize computational efficiency given the large volume of the training set, we estimate the density field using a representative subsample of $N=20,000$ objects and evaluate it on a discrete grid (typically $50^\text{dim}$ bins), performing trilinear interpolation for individual query points.

\par The definition of the feature space $\mathbf{x}$ varies depending on the object class (Galaxy, Star, or QSO) to capture the most relevant physical drivers for each population. Furthermore, to explicitly address the class imbalance present in the training set, we normalize the individual sampling weights such that the total probability mass assigned to each class is equal to $1/3$. This ensures that the model encounters galaxies, stars, and QSOs with equal frequency during the training process, preventing bias toward the most numerous populations.

\subsection{Galaxies}

\par For galaxies, we adopt a two-stage reweighting scheme. First, we compute a probability weight based on physical properties ($w_{\text{phys}}$) to ensure coverage of diverse star formation histories and ionization mechanisms. Second, we compute a weight based on redshift ($w_{z}$) to flatten the redshift distribution. The final training probability is the product of these two components:
\begin{equation}
    w_{\text{total}} = w_{\text{phys}} \times w_{z}.
\end{equation}

\par The physical weight $w_{\text{phys}}$ is determined hierarchically depending on the redshift of the galaxy and the availability of valid measurements for key variables. We segment the galaxy population into three redshift regimes corresponding to the visibility of specific emission lines within the J-PAS filter system: low-$z$ ($z < 0.49$), intermediate-$z$ ($0.49 \le z < 1.01$), and moderate $z$ ($z \ge 1.01$).

\par Within each regime, we prioritize diagnostic diagrams that maximize physical information.
\begin{itemize}
    \item Low-$z$ ($z < 0.49$): The primary space is defined by the BPT diagram variables: $\log([\text{O}\,\textsc{iii}]/\text{H}\beta)$, $\log([\text{N}\,\textsc{ii}]/\text{H}\alpha)$, and $\log(\text{EW}_{\text{H}\alpha})$. This ensures a balanced sampling of Star-Forming, AGN galaxies, and passive galaxies. If specific lines are unavailable (e.g., missing [N\,\textsc{ii}]), we fallback to secondary spaces such as Mass vs. H$\alpha$, or Mass vs. SFR.
    \item Intermediate-$z$ ($0.49 \le z < 1.01$): We utilize the $\log(\text{EW}_{\text{H}\beta})$, $\log([\text{O}\,\textsc{iii}]/\text{H}\beta)$, and Stellar Mass ($\log M_\star$) space.
    \item Moderate-$z$ ($z \ge 1.01$): We rely on Stellar Mass and $\log(\text{EW}_{\text{[OII]}})$.
\end{itemize}

\par For galaxies lacking valid emission line measurements or physical parameters in the training catalogs, we assign weights based on their location in the color-color space defined by $\log(g/r)$ and $\log(i/r)$ (or their synthetic equivalents). To handle non-positive equivalent widths (absorption or noise) in log-space, we replace values $\le 0$ with a randomized Gaussian draw centered at $0.25\,\AA$ prior to density estimation.

\subsection{Stars}

\par For stellar objects, the objective is to uniformly sample the atmospheric parameter space. We define the feature space $\mathbf{x}$ using the effective temperature ($T_{\text{eff}}$), surface gravity ($\log g$), and metallicity ([Fe/H]) derived from the DESI stellar pipeline.

\par Objects with valid spectroscopic parameters are weighted by the inverse density in this 3D space. For stars where these parameters are undefined or unreliable (e.g., low S/N spectra), we default to a 2D color space defined by the flux ratios $\log(g/r)$ and $\log(i/r)$. 

\subsection{Quasars (QSOs)}

\par For QSOs, the primary observational challenge is the redshift-dependent visibility of broad emission lines and the Lyman-$\alpha$ forest. Therefore, we do not reweight based on intrinsic physical properties (such as black hole mass or luminosity) in this version. Instead, we apply a strictly redshift-based weighting, $w_{\text{QSO}} \propto 1/\rho(z)$, to enforce a flat redshift distribution. This ensures that high-redshift quasars, which are critical for cosmological studies but rare in magnitude-limited samples, contribute significantly to the training loss.

\section{Emission line fluxes}\label{app:fluxes}

In order to recover absolute emission line fluxes, \texttt{OJALA} predicts not only the EW but also the continuum flux densities measured at the \ha\ and \hb\ wavelengths. Together with the EW estimations, this allows us to retrieve absolute fluxes not only for \ha\ and \hb, but also for \oiii\ and \nii, under the assumption that the continuum remains roughly constant for closely neighboring lines. 

In Figure~\ref{fig:flux_residuals}, we show the logarithmic ratio between the predicted J-PAS fluxes and the measured DESI spectroscopic fluxes as a function of the DESI EW for these four primary lines. As shown in the figure, the flux predictions exhibit very low systematic bias across the evaluated range. The primary difference among the lines lies in their dispersion, which ranks from lowest to highest scatter as follows: \ha, \hb, \nii, and \oiii. These results are consistent with the EW performance discussed in the main text. 

\par We note that the scatter of the flux predictions does not exhibit a strong dependence on the EW itself, but rather mirrors the underlying distribution of the training set. The model achieves the highest precision in the regions where the bulk of the galaxy population resides. However, we observe a slight underestimation in the high-EW tail for the majority of the lines. This effect is particularly evident for the \oiii\ line, which presents a distribution with a very long tail. Extreme emission values are intrinsically more difficult to predict due to their sparse representation in the training space, leading the model to regress toward the mean of the densely populated regions.

\par These results are similar to those presented by \citet{2025arXiv251121822F}, who employed BB SED fitting using \texttt{CIGALE} to reconstruct both the continuum emission and the subsequent line flux measurements (see Figure 6 of their work; note that they use a $\log$ scale on the y-axis, so our metrics need to be multiplied by 2.3). Although we are able to recover fluxes for emission lines with lower EWs, disentangle \nii\ from \ha, and make predictions at higher redshifts, our model is slightly less accurate for extreme EWs, especially for \oiii. As mentioned earlier, this is due to the sparse representation of these extreme values in the training set, a limitation inherent to our data-driven approach that does not affect traditional methods based on $\chi^2$ minimization.

\par As a further consistency test, we combine the predicted spectral features to derive the intrinsic H$\alpha$ luminosity ($L_{\text{H}\alpha}$), which can be subsequently used to estimate the SFR over the last $\sim 30$ Myr in SF galaxies \citep{1998ApJ...498..541K}. This calculation requires the simultaneous prediction of multiple quantities: the equivalent widths and continuum flux densities of both H$\alpha$ and H$\beta$, together with the galaxy redshift needed to determine the luminosity distance. We compute the extinction-corrected H$\alpha$ luminosity by accounting for dust attenuation through the Balmer decrement, assuming a Case B intrinsic ratio of H$\alpha$/H$\beta = 2.86$.

\begin{figure}[htbp]
   \centering
   \includegraphics[width=\columnwidth]{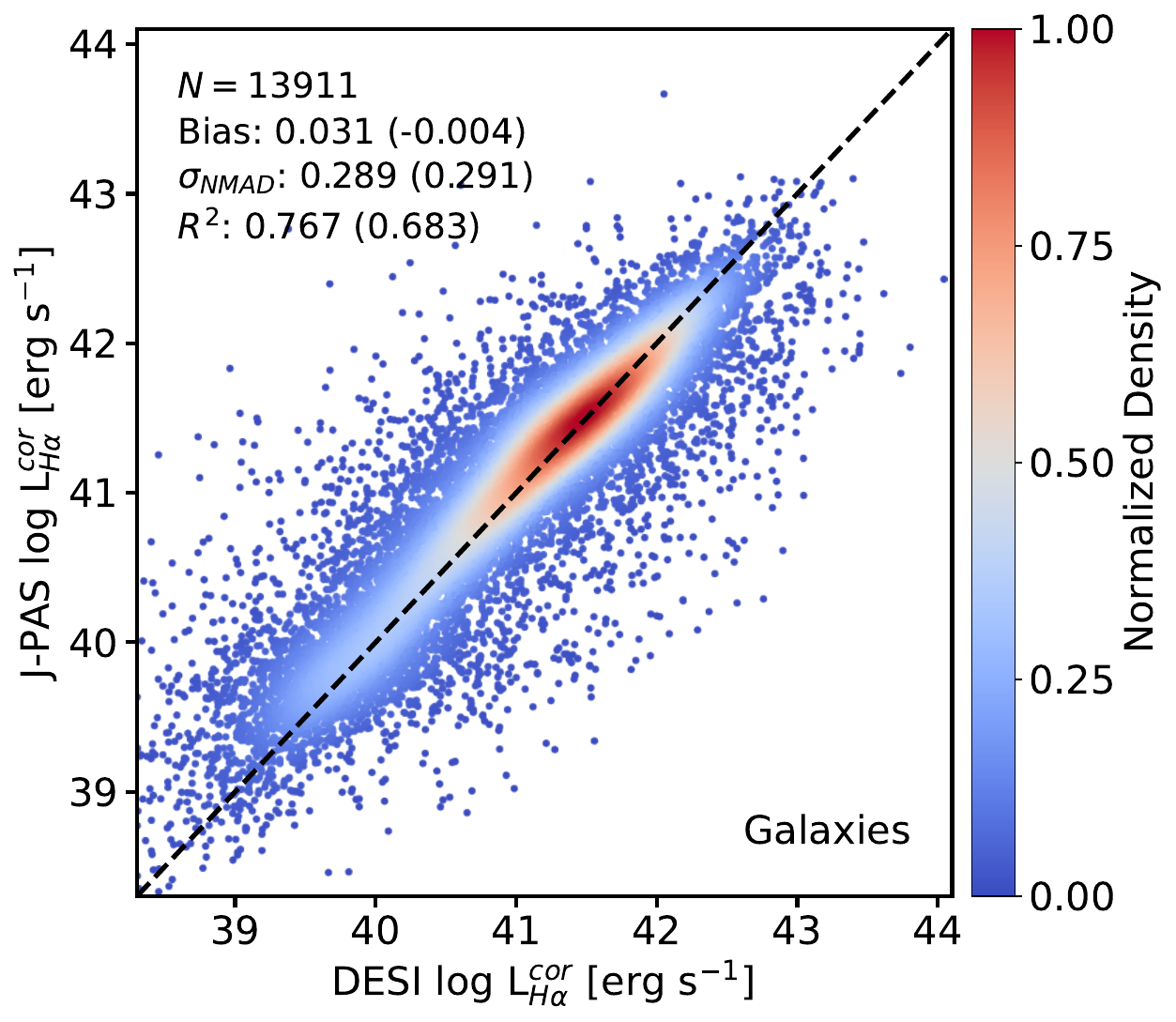}

    \caption{Comparison between the extinction-corrected H$\alpha$ luminosity ($\log L_{\text{H}\alpha}$) derived from DESI spectroscopy and the values inferred from \texttt{OJALA} predictions in \textit{Test-Real}. The calculation accounts for dust extinction using the Balmer decrement derived from the respective H$\alpha$ and H$\beta$ fluxes. The bias, scatter ($\sigma_{\text{NMAD}}$), and $R^2$ are indicated in the legend, where values outside parentheses correspond to \textit{Test-Real} and those in parentheses refer to \textit{Test-Synth}.}

   \label{fig:lum_halpha}
\end{figure}

\par Figure~\ref{fig:lum_halpha} shows the comparison between the intrinsic H$\alpha$ luminosities derived from the model predictions and those obtained from DESI spectroscopy. The agreement is excellent, with minimal bias. We emphasize that no minimum threshold on emission-line intensity is imposed, so the comparison also includes galaxies with weak or negligible emission. For the full sample, we obtain a scatter of $\sigma_{\text{NMAD}} = 0.29$~dex. When restricting the analysis to galaxies with prominent emission, imposing $\mathrm{EW}(\mathrm{H}\alpha) > 10$~\AA, the scatter decreases to $0.24$~dex, demonstrating the higher precision achieved for the actively star-forming population.

\begin{figure*}[htbp]
    \centering
    \includegraphics[width=\textwidth]{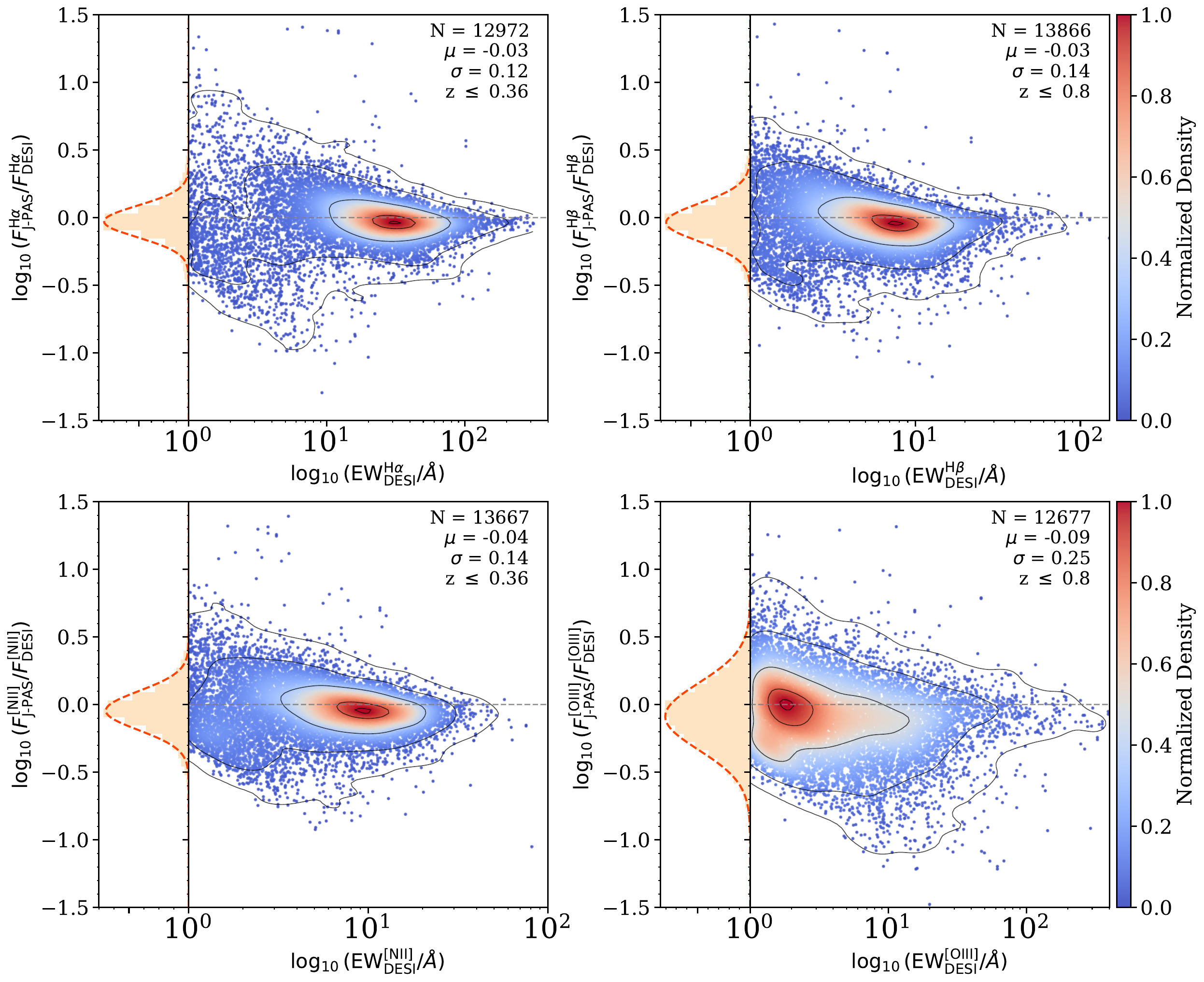}
    \caption{Logarithmic ratio of the predicted J-PAS emission line fluxes on J-PAS real data to the DESI spectroscopic fluxes as a function of the DESI EW for \ha, \hb, \nii, and \oiii. The right sub-panels display the scatter plots colored by normalized density, while the left sub-panels show the corresponding marginal histograms of the residuals along with the best-fit Gaussian distributions. The number of objects ($N$), bias ($\mu$), and dispersion ($\sigma$) are indicated for each line.}
    \label{fig:flux_residuals}
\end{figure*}

\section{Comparison with Previous ANN-based Emission Line Estimators}\label{app:ANN}

\par
\begin{figure*}
    \centering
    \includegraphics[width=\textwidth]{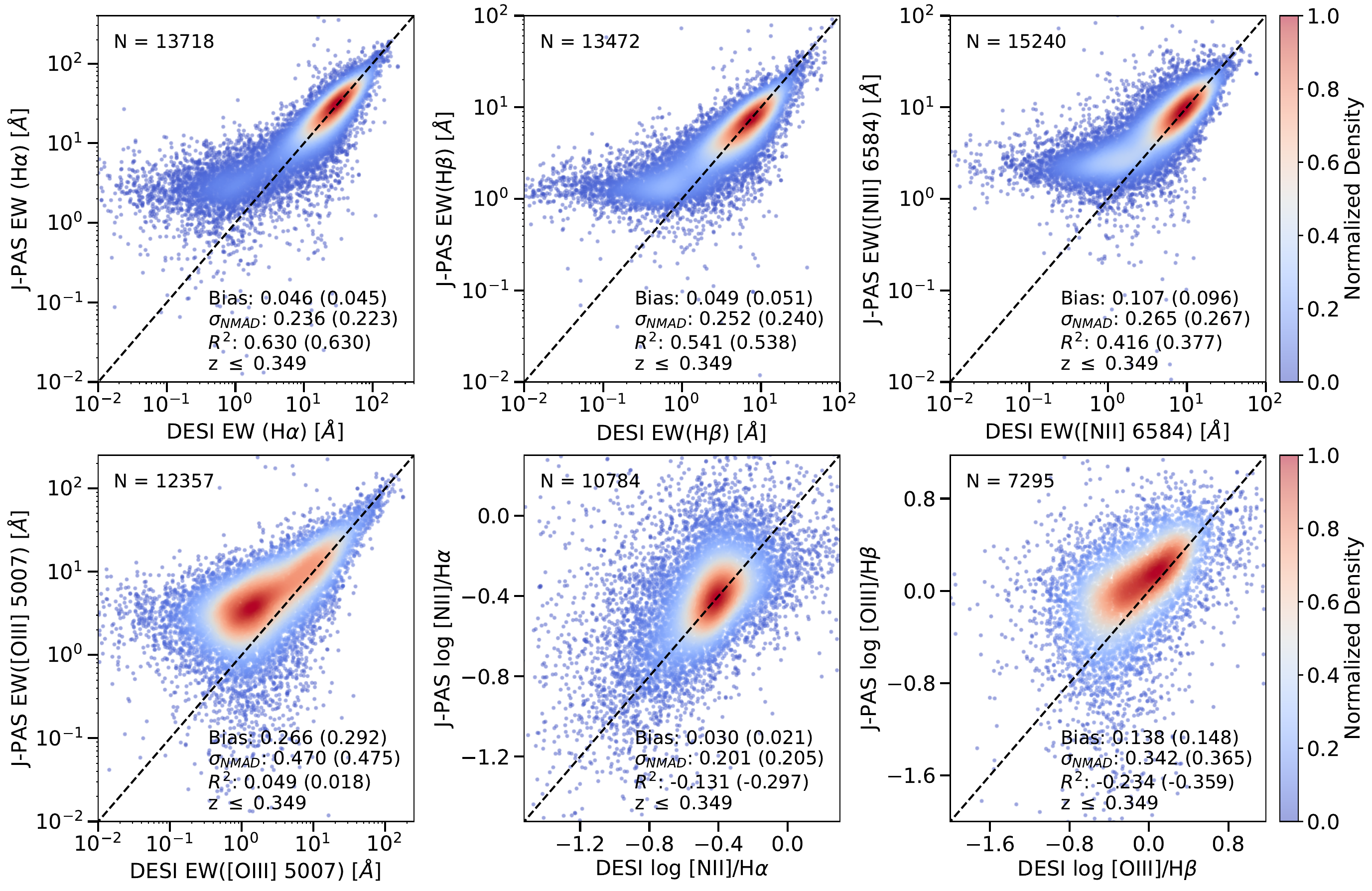}
    \caption{Comparison between DESI spectroscopic EW measurements and the baseline ANN model predictions for J-PAS real data (density plots). Metrics are provided in each panel along with the number of galaxies ($N$). The top row displays the hydrogen Balmer lines and \nii, while the bottom row shows \oiii\ and the diagnostic ratios. Note the use of log scales for EW plots. Statistics include the bias, $\sigma_{\text{NMAD}}$, and $R^2$. Note that the redshift range for this comparison is restricted to $z < 0.35$.}
    \label{fig:EL_scatter_ANN}
\end{figure*}

 In this section, we evaluate the performance of our previous emission line estimator \citep{2021A&A...647A.158M} (hereafter the ANN model) in the context of the J-PAS-DESI cross-match. The performance of the ANN model is presented in Fig.~\ref{fig:EL_scatter_ANN}, following the same format as Fig.~\ref{fig:EL_scatter},while the comparison with the \texttt{OJALA} model (bias, scatter, and $R^2$) for the common set of galaxies is detailed in Table~\ref{tab:model_comparison}.

\begin{table}[h!]
    \centering
    \caption{Performance comparison (Bias, $\sigma_{\text{NMAD}}$, and $R^2$) between the \texttt{OJALA} model and the baseline ANN model for the estimation of EWs and emission-line ratios. The values correspond to the global metrics for the galaxy test sample.}
    \label{tab:model_comparison}
    \resizebox{\columnwidth}{!}{%
    \setlength{\tabcolsep}{5pt}
    \begin{tabular}{lcccccc}
        \toprule
        & \multicolumn{3}{c}{\texttt{OJALA}} 
        & \multicolumn{3}{c}{ANN} \\
        \cmidrule(lr){2-4} \cmidrule(lr){5-7}
        Line / Ratio
        & Bias & $\sigma_{\text{NMAD}}$ & $R^2$
        & Bias & $\sigma_{\text{NMAD}}$ & $R^2$ \\
        \midrule
        $\log_{10}$ EW(H$\alpha$)
        & 0.033 & 0.159 & 0.842
        & 0.046 & 0.236 & 0.630 \\

        $\log_{10}$ EW(H$\beta$)
        & 0.024 & 0.184 & 0.735
        & 0.049 & 0.252 & 0.541 \\

        $\log_{10}$ EW([\text{N\textsc{ii}}] 6584)
        & 0.042 & 0.186 & 0.731
        & 0.107 & 0.265 & 0.416 \\

        $\log_{10}$ EW([\text{O\textsc{iii}}] 5007)
        & 0.018 & 0.282 & 0.660
        & 0.266 & 0.470 & 0.049 \\
        \midrule
        $\log_{10}$([\text{N\textsc{ii}}]/H$\alpha$)
        & 0.020 & 0.109 & 0.648
        & 0.030 & 0.201 & -0.131 \\

        $\log_{10}$([\text{O\textsc{iii}}]/H$\beta$)
        & 0.005 & 0.228 & 0.424
        & 0.138 & 0.342 & -0.234 \\
        \bottomrule
    \end{tabular}%
    }
\end{table}

\par The baseline ANN model consists of artificial neural networks developed using synthetic J-spectra derived from distinct spectroscopic datasets, including IFU data (MaNGA \citep{2015ApJ...798....7B} and CALIFA \citep{2012A&A...538A...8S}) and single-fiber observations (SDSS). This cross-survey strategy was intended to ensure a robust generalization for the final application on J-PAS data. The objective of that model was to predict emission lines for \ha, \hb, \oiii, and \nii\ in the redshift range $0 < z < 0.35$. However, regardless of the specific training configuration, the mapping NB photometry to emission line properties changes dynamically as the spectral features shift across the J-PAS filter system. Given the computational resources and model architecture available at the time, the optimal strategy to manage this complexity involved training multiple, independent ANNs across narrow redshift bins. This segmentation was necessary to simplify the feature space within each interval, implying that the spectroscopic redshift was required as an explicit input feature to select the appropriate network. In contrast, as demonstrated in this work, \texttt{OJALA} learns the global spectral dependencies and does not require prior redshift information to accurately predict emission lines or any other physical property.

\par Furthermore, the ANNs were trained exclusively on spectra with high S/N detections of the target lines, effectively limiting the training sample to SF or active galaxies. This selection was made to circumvent the issue of the heterogeneous presence of quiescent versus SF spectra. Traditional ML approaches typically require fixed input and output vectors; thus, including quiescent spectra would imply handling missing data or zeros in the target variables, which is non-trivial to manage within a standard Mean Squared Error loss function. As a consequence, the ANN predictions tend to saturate prominently at higher EWs in the low EW regime (see Fig.~\ref{fig:EL_scatter_ANN}).

\par The flexibility of the \texttt{OJALA} architecture effectively resolves these limitations. Unlike traditional architectures constrained by fixed input and output vectors, the Transformer processes data as sequences of tokens, allowing for the natural handling of missing photometric bands or emission line measurements without requiring imputation. Moreover, the probability sampling scheme we employ (see Appendix~\ref{app:sampling}) ensures that the model learns the full parameter space of galaxy SEDs effectively. This strategy allows \texttt{OJALA} to properly characterize not only quiescent galaxies, which lack emission lines, but also underrepresented populations such as very metal-poor galaxies or AGN. While these objects were present in the ANN training sets, their statistical influence was diluted in the absence of a targeted sampling strategy, leading to suboptimal generalization in those specific regimes.

\par As shown in Table~\ref{tab:model_comparison}, \texttt{OJALA} significantly outperforms the ANN baseline across all metrics. This superior performance is attributable to a combination of factors. First, \texttt{OJALA} benefits from a much richer training sample based on DESI DR1, which covers a wider redshift range, extends to fainter magnitudes, and comprises approximately 14 million galaxies, compared to the $\sim$300k spectra from the IFU training set used for the ANN. Second, the model architecture is vastly more sophisticated: with 4.6 million parameters (versus $\sim$1000 in the ANN), it uses attention mechanisms to capture subtle, non-local correlations in the data. Finally, the generation of more realistic synthetic data, combined with the implementation of UDA, makes \texttt{OJALA} a more robust, powerful, and versatile model.

\section{The SFMS, comparison with SED fitting}\label{app:SED_fitting}

In this section, we analyze the results of \texttt{OJALA} for the SFR and stellar masses by comparing them with an independent SED fitting approach that has been successfully used in the past for analyzing the stellar population properties of galaxies in miniJPAS \citep[]{2021A&A...649A..79G}. Specifically, we use \texttt{BaySeAGal}, a parametric spectro-photometric fitting code designed to infer stellar population properties of galaxies from multi-band photometry. In this framework, the observed SED, sampled through NB filters, is modeled as the convolution of a parametric star formation history (SFH) with stellar population synthesis models, including the effects of dust attenuation. Specifically, we use the \citep{2003MNRAS.344.1000B} SSP models adopting a  \citep{2003PASP..115..763C} IMF. The intrinsic model spectrum can be written as:

\begin{equation}
L_{\lambda}(\Theta)=e^{-q_{\lambda}\tau_V}\int SSP_{\lambda}(t,Z)\,\psi(t;\Theta)\,dt ,
\end{equation}

where $SSP_{\lambda}(t,Z)$ represents the spectrum of a simple stellar population of age $t$ and metallicity $Z$, $\psi(t;\Theta)$ is the assumed SFH, and $\Theta$ denotes the set of free parameters describing the stellar population, including metallicity, dust attenuation, and the parameters controlling the temporal evolution of star formation. In this analysis, we adopt delayed-$\tau$ star formation histories:
\begin{equation}
\psi(t)=k\,\frac{(t_0-t)}{\tau} e^{-(t_0-t)/\tau},
\end{equation}

where $t_0$ is the onset of star formation and $\tau$ defines the characteristic timescale of the star formation process.

To constrain the model parameters, \texttt{BaySeAGal} compares the synthetic magnitudes predicted by the model with the observed photometric magnitudes measured across the full set of filters. The redshift is fixed to the \texttt{LePHARE} estimate for each galaxy, and filters strongly affected by emission lines are excluded to ensure that the fit is primarily driven by the stellar continuum. \texttt{BaySeAGal} adopts a Bayesian framework based on Markov Chain Monte Carlo sampling to explore the parameter space and estimate the posterior probability distributions of the physical parameters. This approach provides both best-fit estimates and uncertainties for quantities such as stellar mass, stellar age, metallicity, dust attenuation, and intrinsic colors. The SFR is estimated by integrating the SFH over the last 100 Myr. 

We run \texttt{BaySeAGal} on the J-PAS--DESI cross-match using the \texttt {AUTO}  photometry, which ensures all the light of the galaxy contributes to the SED. Unlike the \texttt{CIGALE} reference catalog used for training, this fit does not use the BB photometry from the DESI Legacy Survey or WISE; instead, it relies exclusively on the J-PAS NB filters together with the $i_{\text{SDSS}}$ reference band. 

\begin{figure}[htbp]
    \centering
    \includegraphics[width=\columnwidth]{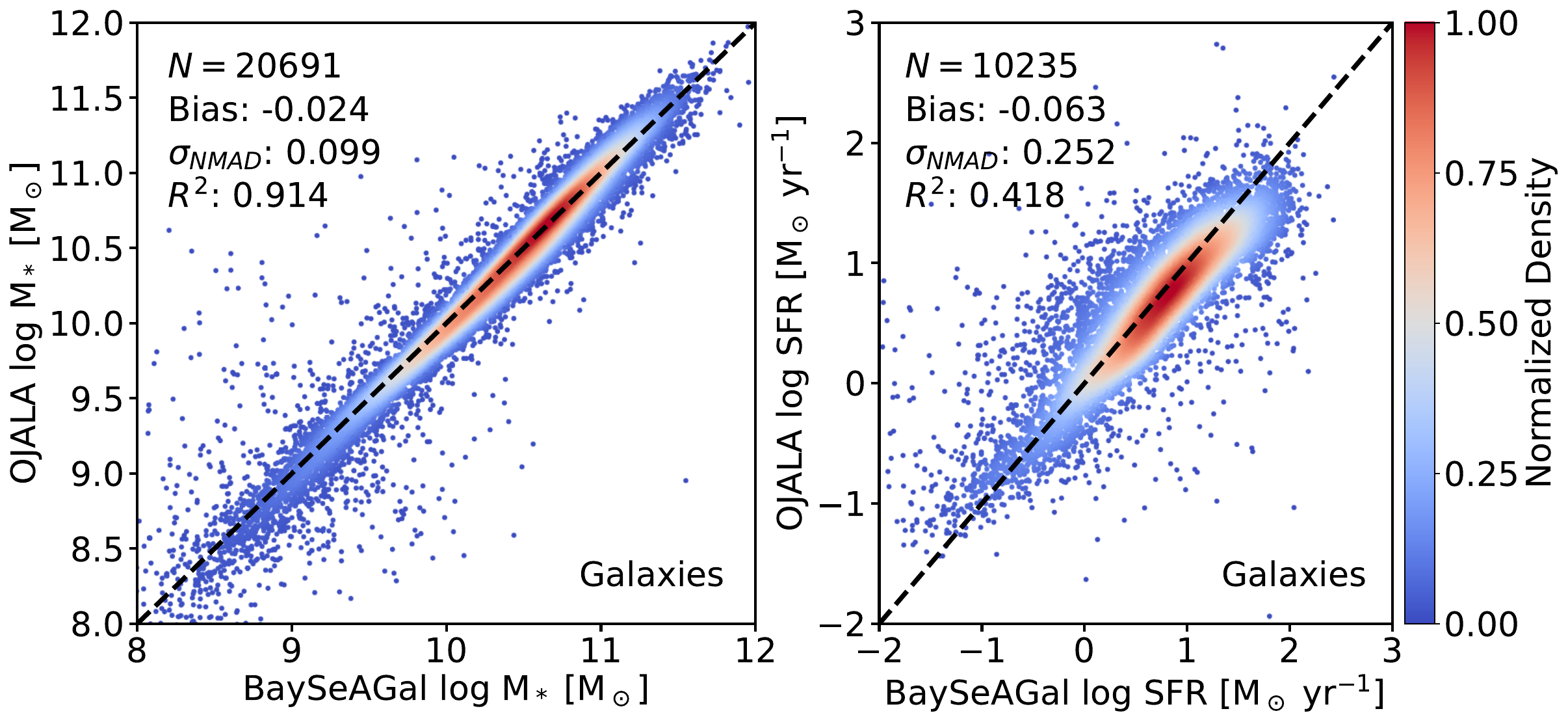}
    \caption{Comparison between the predicted stellar masses (left panel) and star formation rates (right panel) from \texttt{BaySeAGal} and \texttt{OJALA} for the J-PAS--DESI cross-match sample. The points are color-coded by normalized density. The number of objects ($N$), bias (median offset), scatter ($\sigma_{\text{NMAD}}$), and $R^2$ are indicated in the top left of each panel. The dashed black line represents the one-to-one relation.}
    \label{fig:LOGM_LOGSFR_Bay}
\end{figure}

In Figure \ref{fig:LOGM_LOGSFR_Bay}, we compare the predicted stellar masses and SFRs from \texttt{OJALA} and \texttt{BaySeAGal}, applying the same quality cuts used in Fig.~\ref{fig:logM_logSFR}.As observed, the agreement is excellent. For the stellar mass, we obtain a bias of $-0.024$ dex and a scatter of $\sigma_{\text{NMAD}} = 0.099$ dex, with $R^2 = 0.914$. For the SFR, the bias is $-0.063$ dex with a scatter of $\sigma_{\text{NMAD}} = 0.252$ dex and $R^2 = 0.418$. These metrics are very similar to those obtained when comparing \texttt{OJALA} predictions with the \texttt{CIGALE} DESI VAC. 

This consistency is remarkable when contextualizing the different underlying assumptions of both SED fitting codes. \texttt{OJALA} was trained to reproduce \texttt{CIGALE} estimates, which are derived primarily from BB photometry and rely on specific intrinsic assumptions. For instance, the \texttt{CIGALE} catalog models the SFH with an additional late starburst component, parameterized as $\text{SFR}(t) = \text{SFR}_{\text{delayed}}(t) + \text{SFR}_{\text{burst}}(t)$. Furthermore, despite fitting BB photometry, \texttt{CIGALE} explicitly incorporates nebular emission and an AGN contribution, even though the present comparison is restricted to the galaxy sample. 

\begin{figure*}[htbp]
    \centering
    \includegraphics[width=\textwidth]{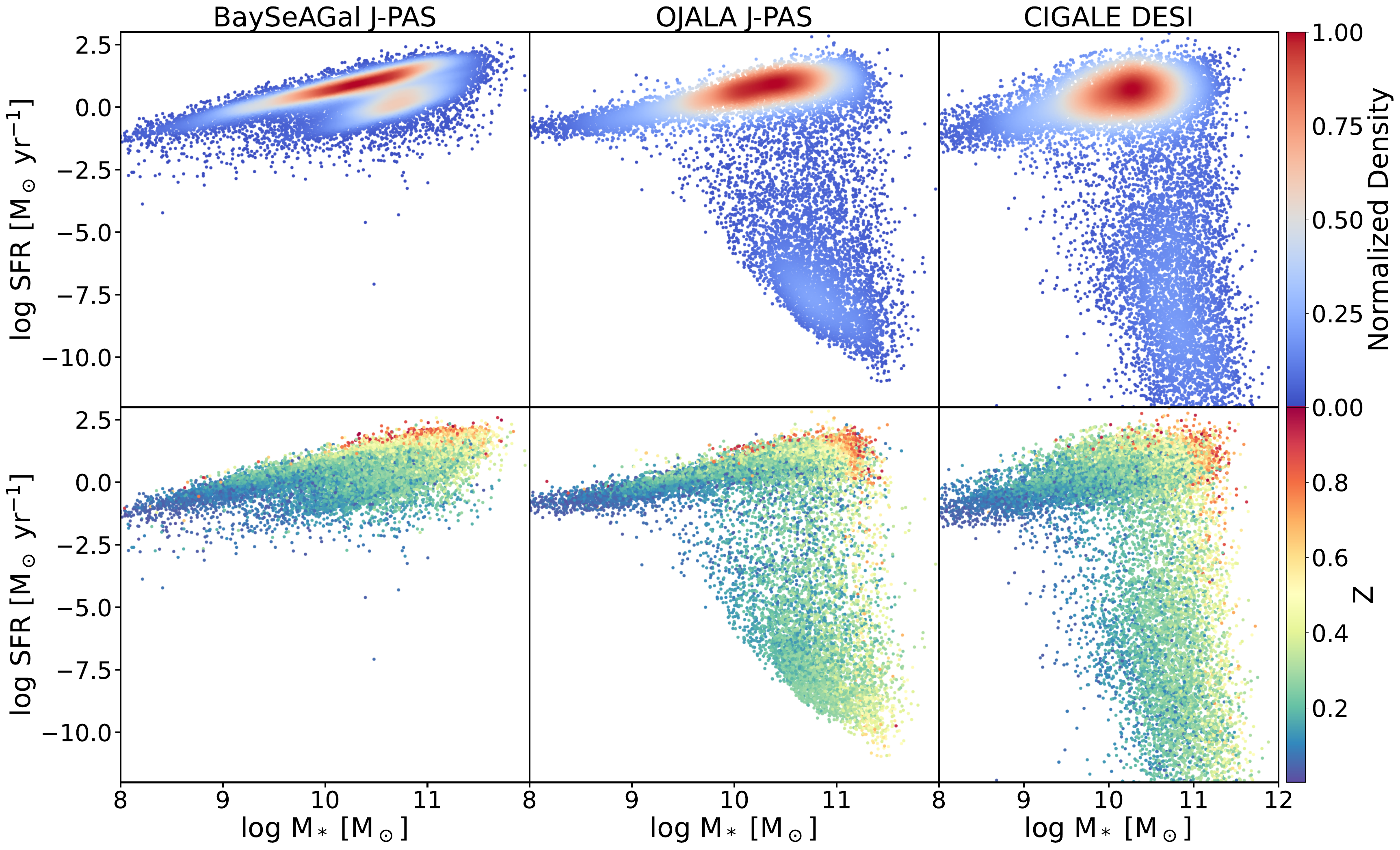}
    \caption{Star Formation Main Sequence (SFMS) derived for all galaxies in the J-PAS--DESI cross-match at $z < 1$. The columns display the distributions inferred by \texttt{BaySeAGal} (left), \texttt{OJALA} (middle), and \texttt{CIGALE} (right). The top row is color-coded by normalized density, highlighting the core of the main sequence and the distribution of quiescent galaxies. The bottom row is color-coded by spectroscopic redshift ($Z$), illustrating the expected trend of increasing SFR at higher redshifts across all three methodologies.}
    \label{fig:SFMS_combined}
\end{figure*}

Finally, in Figure \ref{fig:SFMS_combined}, we present the Star Formation Main Sequence (SFMS) derived for all galaxies in the J-PAS--DESI cross-match at $z < 1$, comparing the distributions from \texttt{BaySeAGal}, \texttt{OJALA}, and \texttt{CIGALE}. A remarkable agreement is observed for star-forming galaxies located on the main sequence. However, expected differences emerge for quiescent galaxies that fall below the SFMS. While \texttt{BaySeAGal} produces a distinct, concentrated cluster for high-mass galaxies leaving the main sequence, \texttt{CIGALE} (and consequently \texttt{OJALA}) distributes this population across a wider range of lower SFR values. This low specific SFR regime is characterized by high uncertainties, and these galaxies were  filtered out in previous direct comparisons. In this regime, the optical SED lacks sufficient information for stellar population fits to accurately distinguish residual star formation rates, leading to the different clustering behaviors observed between the codes. Additionally, the color gradient in the bottom panels clearly illustrates how the SFR consistently increases for higher redshifts, a trend successfully captured by all three methods.

\end{document}